\def\mass#1{${\mathrm{#1\, M}_\odot}$}
\def\mmass#1{{\mathrm{#1\, M}_\odot}}

\def\chem#1#2{$\mathrm{^{#2}\kern-0.8pt#1}$}
\def\mchem#1#2{\mathrm{^{#2}\kern-0.8pt#1}}
\def\reac#1#2#3#4#5#6{$\mathrm{\, ^{#2}\kern-0.8pt{#1}\, ({#3}\, ,{#4})\, {}^{#6}\kern-0.8pt{#5}\, }$}
\def\betap#1#2#3#4{$\mathrm{\, ^{#2}\kern-0.8pt{#1}\, (\beta^+)\, {}^{#4}\kern-0.8pt{#3}\, }$}
\def\betam#1#2#3#4{$\mathrm{\, ^{#2}\kern-0.8pt{#1}\, (\beta^-)\, {}^{#4}\kern-0.8pt{#3}\, }$}
\def\reacbp#1#2#3#4#5#6#7#8{$\mathrm{\, ^{#2}\kern-0.8pt{#1}\, ({#3}\, ,{#4})\, {}^{#6}\kern-0.8pt{#5}\, (\beta^+)\, {}^{#8}\kern-0.8pt{#7}\, }$}
\def\reacbm#1#2#3#4#5#6#7#8{$\mathrm{\, ^{#2}\kern-0.8pt{#1}\, ({#3}\, ,{#4})\, {}^{#6}\kern-0.8pt{#5}\, (\beta^-)\, {}^{#8}\kern-0.8pt{#7}\, }$}
\def\simgr{\mathbin{\;\raise1pt\hbox{$>$}\kern-8pt\lower3pt\hbox{$\sim$}\;}}
\def\simlr{\mathbin{\;\raise1pt\hbox{$<$}\kern-8pt\lower3pt\hbox{$\sim$}\;}}

\documentclass{aa}
\usepackage{graphics}
\begin{document}

\thesaurus{06(08.16.4; 08.03.1; 08.05.3; 08.09.3; 08.01.1)}

\title{Sodium production in asymptotic giant branch stars}
 
\author{Nami Mowlavi}
                                

\institute{Geneva Observatory, CH-1290 Sauverny, Switzerland}

\date{Received 20 May 1999 / Accepted 23 July 1999}

\maketitle

\begin {abstract}
  A new scenario is presented for the production of \chem{Na}{23} in asymptotic
giant branch (AGB) stars. The scenario takes advantage of the periodic third
dredge-up episodes
characterizing those stars, which mix primary
\chem{C}{12} from their intershell layers to their surface. Two
successive interpulse/pulse/dredge-up sequences
are then required to produce \chem{Na}{23}.
During the first sequence carbon and oxygen are converted into \chem{N}{14}
by the hydrogen burning shell, and subsequently transformed into \chem{Ne}{22}
by the helium burning shell.
During the second sequence, \chem{Ne}{22} is converted into \chem{Na}{23} by
the hydrogen burning shell, which is brought to the surface by the subsequent
dredge-up episode.  The \chem{Na}{23} produced by this scenario is thus primary.

  The efficiency of this scenario is analyzed
through standard evolutionary AGB model predictions combined with synthetic
calculations for the surface chemical evolution. It is shown that primary
\chem{Na}{23} can efficiently be produced as soon as the surface C+N+O abundance
enhancement reaches a certain level depending on the stellar metallicity. The
required surface C+N+O abundance enhancement amounts to $\sim$0.4~dex in solar metallicity stars,
and to $\sim$0.8~dex at a metallicity five times less than solar.

  An {\it analytical} study of Na production further reveals that the surface
\chem{Na}{23} abundance asymptotically evolves to a 
`line of primary sodium enrichment' (LOPSE) in the [C+N+O]$-$[\chem{Na}{23}] diagram.
That LOPSE represents the \chem{Na}{23} abundance evolution predicted in
zero metallicity AGB stars experiencing third dredge-up episodes.
An analytical relation for the surface
\chem{Na}{23} abundance evolution as a function of the surface C+N+O abundance is
provided.

The predicted surface \chem{Na}{23} enhancements can exceed 0.5~dex depending on the
level of surface \chem{C}{12} enrichment, and increases with decreasing stellar
metallicity. The quantitative prediction of \chem{Na}{23} surface abundances,
however, is presently subject to a high level of uncertainty, partly due
to the still poor quantitative prediction of the
structural evolution of AGB stars (dredge-up episodes in particular),
and partly due to the uncertainties still
affecting some nuclear reaction rates (such as \chem{Na}{23} destruction by
proton capture).

The case of massive AGB stars in which hot bottom burning occurs
is also discussed. The production of secondary sodium in those stars is a
natural consequence of \chem{Ne}{22} burning in their envelope, if the
temperature at the base of the envelope exceeds 70 million~K. It requires,
however, many interpulses to be significant. The production of primary sodium
from the dredge-up of primary \chem{Ne}{22} and its subsequent burning in the
envelope, on the other hand, is estimated not to be very efficient,
expect maybe in low-metallicity stars.

An eventual detection of high Na overabundances in carbon stars or
related objects would support the scenario of primary sodium
production in AGB stars. Such an observational evidence may have been found in at
least one post-AGB star.
Further observations of those objects are called for.
Observations of \chem{Na}{23} in planetary nebulae are also encouraged. Finally,
the production of primary \chem{Na}{23} by AGB stars, if confirmed observationally,
may have played a non-negligible role in the chemical evolution of our Galaxy.

\end{abstract}

\section{Introduction}
\label{Sect:introduction}

  Sodium overabundances have been observed in A-F supergiants,
with 0.0$\simlr$[Na/H]$\simlr$0.5 for the F supergiants
($\mathrm{[X]\equiv \log X_{star}-\log X_\odot}$) and
0.7$\simlr$[Na/H]$\simlr$0.8 for the A-type supergiants
(Takeda \& Takada-Hidai \cite{Takeda_Takada-Hidai94} and references
therein).
The presence of enhanced sodium abundances at the surface of those stars is
explained by the operation of the Ne-Na mode of H-burning during the main
sequence phase, which leads to an increase of \chem{Na}{23} in the deep stellar
layers by a factor up to ten. The penetration in those
layers of the convective envelope when the star becomes
a red giant, process called `first dredge-up',
then brings the synthesized sodium to the surface.
First dredge-up predictions indeed confirm the production of
\chem{Na}{23} to the level observed in A-F supergiants
(El Eid \& Champagne \cite{ElEid_Champagne95}).

High \chem{Na}{23} overabundances have also been measured at the surface of red
giants in globular clusters as early as in the late 70's (see Mowlavi
\cite{Mowlavi98} for references).
Again, the Ne-Na chain of H-burning and the operation of first dredge-up are made
responsible for those surface overabundances. The observed overabundances, however,
are higher than those predicted by standard\footnote{Standard models refer, in
this paper, to models using the Schwarzschild criterion to delimit convective
zones without applying any extra-mixing procedure such as overshooting or
diffusive mixing induced by rotation} model predictions, but can be explained if
extra-mixing is assumed to operate below the convective envelope, probably due
to rotationally induced meridional circulation.
In both globular clusters' red giants and near solar metallicity supergiants,
the production of \chem{Na}{23} results from the
transformation of the {\it initial} \chem{Ne}{22}. The sodium
produced in those stars is thus secondary\footnote{
An element is said to be secondary if its synthesis requires the presence of
some elements heavier than helium in the initial stellar composition. In
contrast, an element is said to be primary if it can be synthesized in a star of
the first generation with zero metallicity (the metallicity being the mass
fraction of all elements heavier than helium).}.

In this paper, a new scenario is proposed to produce {\it primary} sodium in
asymptotic giant branch (AGB) stars. The scenario, explained in more details in
Sect.~\ref{Sect:AGB}, takes advantage of the \chem{C}{12} and
\chem{Ne}{22} production in the helium burning shell (HeBS) of those
stars, and of the periodic mixing of those nuclei to the stellar surface. Part
of the \chem{Ne}{22} may then subsequently transform into \chem{Na}{23} in
the hydrogen burning shell (HBS) of those stars. Such a scenario is expected to
operate in carbon stars.

Several standard model stars are computed from their pre-main sequence up to the
AGB phase. The stellar masses and metallicities covered in those
calculations range from $M=1.5$ to \mass{6} and from $Z=0.004$ to 0.02, respectively.
The properties of those standard models are presented in
Sect.~\ref{Sect:standard models}. In particular, the surface
\chem{Na}{23} abundance evolution prior to the AGB phase, and the
abundances of \chem{C}{12}, \chem{Ne}{22} and \chem{Na}{23} predicted in the
intershell layers during the AGB phase are discussed.
Primary sodium production is then studied in detail in Sect.~\ref{Sect:Na23}
using synthetic AGB calculations. The case of the \mass{6} star, which undergoes
hot bottom burning (HBB), is considered separately in Sect.~\ref{Sect:HBB}.
Finally, the observational implications/confirmations of the scenario are briefly
discussed in Sect.~\ref{Sect:observations}.
Conclusions are drawn in Sect.~\ref{Sect:conclusions}.

Two Appendixes are further presented at the end of the paper.
Appendix~A describes
briefly the CNO and Ne-Na modes of hydrogen burning, and presents the yields of
those elements at hydrogen exhaustion as a function of the
H-burning temperature. In particular, the impact of the new
NACRE reaction rates on the \chem{Na}{23} production is discussed.
Appendix~B, on the other hand, presents an analytical study of \chem{Na}{23} production
in AGB stars.

\begin{figure}[t]
  \resizebox{\hsize}{!}{\includegraphics{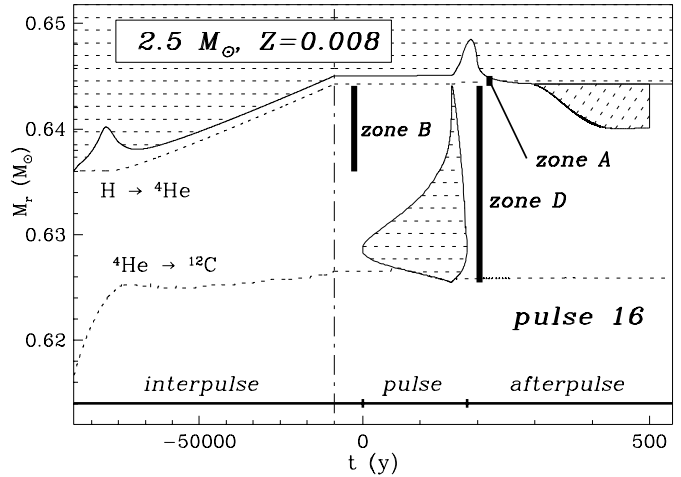}}
  \caption{Structural evolution of the intershell regions of the \mass{2.5} model
           star with $Z=0.008$ during the 15th interpulse and 16th pulse. The origin of the
           abscissa is arbitrarily set to the time of onset of pulse 16. Filled
           regions correspond to convective zones. The dotted lines identify the
           location of maximum energy production in the H-burning (top) and
           He-burning (bottom) layers. The different phases of an instability
           cycle and zones {\it A}, {\it B} and {\it D} in the intershell layers
           described in Sect.~\ref{Sect:AGB} of the main text
           are also indicated. The simulation of a third
           dredge-up, as predicted by models using the prescription of Mowlavi
           (\cite{Mowlavi99}), is shown in the afterpulse phase
           (region hatched at 45 degrees in the right subpanel).
           Note that the time abscissa
           has different scales in the two subpanels displaying
           interpulse 15 and pulse 16.
          }
  \label{Fig:phases}
\end{figure}

\section{AGB stars and \chem{Na}{23} production}
\label{Sect:AGB}

\subsection{AGB stars}
\label{Sect:intershell layers}
 
An AGB star is characterized by an electron degenerate C--O core of 0.5 to \mass{1.2},
by a thin HeBS of about \mass{10^{-2}} capped by a thin HBS of a few \mass{10^{-4}},
and by a deep convective envelope extending from above the HBS up to the surface.
The HeBS is known from numerical simulations to be thermally unstable
and to liberate, periodically and on a short time-scale (of several tens of years),
$10^2$ to $10^6$ times the energy provided by the H-burning shell.
These outbursts lead to the development of a pulse convective tongue in the HeBS.
Figure~\ref{Fig:phases}, for example, displays the structural evolution of the
intershell layers of a \mass{2.5}, $Z=0.018$ model star between its 15th and
16th pulse.

  A thermal instability cycle is divided, for our purposes, into three
phases (illustrated in Fig.~\ref{Fig:phases}): the interpulse, the pulse
and the afterpulse phases. The main characteristics of each of these phases
in relation with sodium production are described as follow (a more detailed
description can be found in Mowlavi \& Meynet, in preparation, hereafter
called MM99):

\paragraph{Interpulse phase:}

 During the interpulse phase, hydrogen burns in a thin layer located
at mass fractions increasing with time, while the HeBS is almost extinct.
The ashes left over by that
HBS accumulate in the underlying He-rich zone (called {\it zone
B}, see Fig.~\ref{Fig:phases}).
Among them, \chem{N}{14} is the main product emerging from the CNO mode of
hydrogen burning, while \chem{Na}{23} is produced in the NeNa chain (see Appendix~A).

\paragraph{Pulse phase:}

As the mass of the He-rich zone increases, a thermal
instability is triggered in the HeBS, and a pulse convective tongue
develops. The pulse starts at the location of maximum
energy production in the HeBS, and extends outward up to close the HBS. As it
grows, it engulfs the ashes of H-burning available in zone {\it B}.
\chem{N}{14}, in
particular, is mixed down to the high temperatures characterizing the pulse, and
burns through $\alpha$-capture into \chem{O}{18}. This nuclide, in turn, may burn
and synthesize \chem{Ne}{22} through \reac{O}{18}{\alpha}{\gamma}{Ne}{22}.
Those products are left over by the pulse in zone {\it D}, together with
primary \chem{C}{12} synthesized by the 3-$\alpha$ reaction.

\paragraph{Afterpulse phase:}

During the afterpulse phase, the structural evolution of the
intershell layers is dominated by thermal relaxations. The temperature in those
intershell layers drop, and the convective envelope deepens into the
H-depleted and, eventually, into the C-rich layers.
The ashes of H-burning which escape the ingestion by the pulse convective tongue
(zone {\it A} in Fig.~\ref{Fig:phases}), and part of the material of zone {\it D} processed by
the pulse (containing in particular \chem{C}{12}, \chem{Ne}{22} and \chem{Na}{23})
are mixed to the surface.
This process of mixing C-rich material into the envelope is called
'third dredge-up' (3DUP).

\subsection{Sodium production}
\label{Sect:Na23scenario}

  Sodium is produced in the HBS through \reac{Ne}{22}{p}{\gamma}{Na}{23}. This
reaction essentially transforms into \chem{Na}{23} all the \chem{Ne}{22} supplied
by the envelope to the HBS. For solar initial composition
(displayed in Table~\ref{Tab:initial abundances}), it would lead to an intershell
\chem{Na}{23} mass fraction of $1.61 \times 10^{-4}$.
At temperatures exceeding $35\times 10^6$~K, the slight burning of the
more abundant \chem{Ne}{20} nuclei contributes to an
extra-production of \chem{Na}{23} (maximum 60\% at 55-$60\times 10^6$~K, see
Fig.~\ref{Fig:NeNayields} of Appendix~A). Above $60\times 10^6$~K,
however, \chem{Na}{23} may be destroyed by p-capture through
\reac{Na}{23}{p}{\alpha}{Ne}{20} and/or \reac{Na}{23}{p}{\gamma}{Mg}{24}.
A more detailed description of the NeNa chain is provided in Appendix~A.
The actual predictions from standard AGB model
calculations are discussed in Sect.~\ref{Sect:Na23standard}.

\begin{table}
\caption[]{\label{Tab:initial abundances}
           Solar abundances, in mass fraction, of the most abundant isotopes of
           the C, N and O elements, and of the stable nuclei
           involved in the Ne-Na chain (the solar metallicity is taken
           equal to $Z=0.018$).
          }
\begin{tabular}{c c}
\hline
\noalign{\smallskip}
 nuclei & solar mass fraction\\
\noalign{\smallskip}
\hline
\noalign{\smallskip}
\chem{C}{12}  & $2.878\times 10^{-3}$\\
\chem{N}{14}  & $1.049\times 10^{-3}$\\
\chem{O}{16}  & $9.103\times 10^{-3}$\\
\chem{Ne}{20} & $1.536\times 10^{-3}$\\
\chem{Ne}{21} & $3.925\times 10^{-6}$\\
\chem{Ne}{22} & $1.236\times 10^{-4}$\\
\chem{Na}{23} & $3.166\times 10^{-5}$\\
\noalign{\smallskip}
\hline
\end{tabular}
\end{table}

It must be noted that the impact of that intershell secondary
\chem{Na}{23} on the surface sodium abundance is {\it negligible}, though.
This is due to the fact that the mass of intershell material dredged-up to the
surface is small (\mass{<0.01}, see Fig.~\ref{Fig:phases}) compared to
the mass contained in the envelope (\mass{\simgr 1}).

If we consider, however, the \chem{C}{12} and \chem{Ne}{22} dredged-up from the HeBS into
the envelope, and the resulting nucleosynthesis during the
interpulse and pulse phases, then a much larger amount
of \chem{Na}{23} may be synthesized. The chain of events is summarized as follow:

\vskip 1mm
\noindent 1) 3DUP mixes into the envelope some primary \chem{C}{12} produced
in the intershell layers by the 3-$\alpha$ reaction;

\vskip 1mm
\noindent 2) during the following interpulse, the \chem{C}{12} (and \chem{O}{16})
provided by the envelope to the HBS is burned into \chem{N}{14} through the CNO
cycle, which accumulates in zone $B$ of the intershell layers;

\vskip 1mm
\noindent 3) the \chem{N}{14} left over in zone {\it B} is engulfed in the
next pulse, and contributes to the nucleosynthesis occurring in that pulse. In
particular, \chem{Ne}{22} is synthesized through
\reac{N}{14}{\alpha}{\gamma}{F}{18}($\beta^+$)\reac{O}{18}{\alpha}{\gamma}{Ne}{22};

\vskip 1mm
\noindent 4) part of that \chem{Ne}{22} emerging from the pulse is dredged-up in
the envelope during the afterpulse phase;

\vskip 1mm
\noindent 5) the \chem{Ne}{22} provided by the envelope to the HBS during the
next interpulse is transformed into \chem{Na}{23} by the NeNa chain. The fraction
of that \chem{Na}{23} which survives destruction by p-capture accumulates in
zone $B$ of the intershell layers;

\vskip 1mm
\noindent 6) the \chem{Na}{23} left over in zone {\it B} is engulfed in the
next pulse, but is unaffected by He-burning nucleosynthesis. It thus emerges
from the pulse unburned but diluted over zone {\it D}, and is partly mixed
to the surface by the following 3DUP.

\vskip 1mm
  In summary, the sodium produced in this chain of events essentially
results from the conversion of the envelope's CNO into \chem{Ne}{22}
during an interpulse/pulse/3DUP chain of events, followed by the
conversion of \chem{Ne}{22} into \chem{Na}{23} during a second
interpulse/pulse/3DUP chain of events. Two dilution processes are thus
necessary. Assuming a solar repartition of the elements heavier than
helium, it can be shown (Sect.~\ref{Sect:Na23predictions}) that the
{\it initial} CNO abundances are insufficient to produce \chem{Na}{23} in
excess of its initial abundance. The dredge-up of primary \chem{C}{12}
from the HeBS is thus essential to increase the envelope's C+N+O abundance,
and to lead to significant \chem{Na}{23}
production. The sodium produced in this way is thus primary.

\section{Standard models}
\label{Sect:standard models}

\subsection{The models}
\label{Sect:models}

\begin{table}
\caption[]{\label{Tab:12DUP}
           \chem{Na}{23} surface mass fraction predicted
           after first dredge-up ($X_1$) and at the end 
           of the E-AGB phase ($X_2$). The initial mass fraction
           $X_{init}$ is also given for comparison.
           The mass fractions are given in units of $10^{-5}$.
           The increase of the surface \chem{Na}{23} at the onset
           of the first pulse relative to its initial abundance
           is given in the fifth (in number ratio) and sixth
           (in dex) columns.
          }
\begin{tabular}{cccccccc}
\hline
\noalign{\smallskip}
 model star       & & $X_{init}$     & $X_1$     & $X_2$     & & \multicolumn{2}{c}{$X_2/X_{init}$}\\
 $M/M_{\odot}, Z$ & & $10^{-5}$ & $10^{-5}$ & $10^{-5}$ & & ratio & dex \\
\noalign{\smallskip}
\hline
\noalign{\smallskip}
 6.0, 0.020 & & 3.52 & 5.38 & 6.32 & & 1.80 & 0.25 \\
 4.0, 0.018 & & 3.17 & 4.82 & 4.88 & & 1.54 & 0.19 \\
 3.0, 0.020 & & 3.52 & 5.20 & 5.20 & & 1.48 & 0.17 \\
 2.5, 0.018 & & 3.17 & 4.54 & 4.54 & & 1.43 & 0.16 \\
 1.5, 0.018 & & 3.17 & 3.37 & 3.37 & & 1.06 & 0.03 \\
\noalign{\smallskip}
 2.5, 0.008 & & 1.41 & 2.17 & 2.17 & & 1.54 & 0.19 \\
 2.5, 0.004 & & 0.70 & 1.15 & 1.16 & & 1.66 & 0.22 \\
\noalign{\smallskip}
\hline
\end{tabular}
\end{table}

  Seven model stars are evolved from the pre-main sequence phase up to
the 15-30th pulse in the AGB phase. The initial masses $M$ and
metallicities $Z$ are ($M$/\mass{},~$Z$)$\;=\;$(1.5,~0.018),
(2.5,~0.018), (3,~0.020), (4,~0.018), (6,~0.020), (2.5,~0.008)
and (2.5,~0.004).

The stellar evolution code originates from Mowlavi
(\cite{Mowlavi95}, see also MM99 for more details about the models).
It was designed to follow as accurately as possible
the structural and chemical evolution of AGB stars.
The abundance profiles of 47 nuclei from H to S, linked by a nuclear reaction
network of 160 reactions, are followed consistently with the stellar structure
all along the stars evolution. The initial mass fractions of hydrogen and helium are
taken to be $X=1-Y-Z$ and $Y=0.23+2.5\times Z$, respectively. The abundances of
the other elements are taken from Anders \& Grevesse (\cite{Anders_Grevesse89})
scaled to the required metallicity. In particular, the solar abundances, taken
at $Z=0.018$, of those nuclei involved in
the synthesis of \chem{Na}{23} are listed in Table~\ref{Tab:initial abundances}.
The rates of the nuclear reactions involved in the Ne-Na chain are given in
Appendix~A. Recently, a new compilation of reaction rates has been made available
by the NACRE project. Their impact on the production of \chem{Na}{23} in AGB
stars is also discussed in that Appendix.

The convective zones are delimited by the standard Schwarzschild criterion, with
no extra-mixing beyond the convective borders so defined. Such standard models
have been shown to fail to consistently reproduce the 3DUP phenomenon (see
Mowlavi \cite{Mowlavi99} and references therein). The use of some sort of extra-mixing
is required to model the 3DUP episodes. Such procedures,
however, are very computer time consuming. For that reason, and in order to
analyze the sodium production in a variety of AGB models of different masses
and metallicities, the models presented in this section are performed without
extra-mixing. The surface \chem{Na}{23} abundances are then computed a posteriori
using synthetic AGB calculations. This is done in
Sect.~\ref{Sect:Na23}.

\subsection{Surface $^{23}Na$ abundances prior to the thermally pulsing AGB phase}
\label{Sect:12DUP}

In standard models not experiencing 3DUP episodes during the AGB phase, the
surface \chem{Na}{23} abundance of red giants can only be affected by the first
and, possibly, second\footnote{
The second dredge-up occurs at the end of the early AGB phase (i.e. before the
thermally pulsing AGB phase) and only for stars more massive than \mass{\sim 4} at
$Z=0.02$ (\mass{\sim 2.5} at $Z=0.004$).
}
dredge-ups. The surface sodium abundances predicted after those
dredge-up episodes are shown in Table~\ref{Tab:12DUP} for all our stars.
The \chem{Na}{23} abundance enhancements reach 40 to 80\% (0.15-0.25~dex) relative
to its initial abundance in all the intermediate-mass stars considered in this paper,
and less than 10~\% in the \mass{1.5} star. Those predictions agree with
the ones available in the literature.

\begin{figure}
  \resizebox{\hsize}{!}{\includegraphics{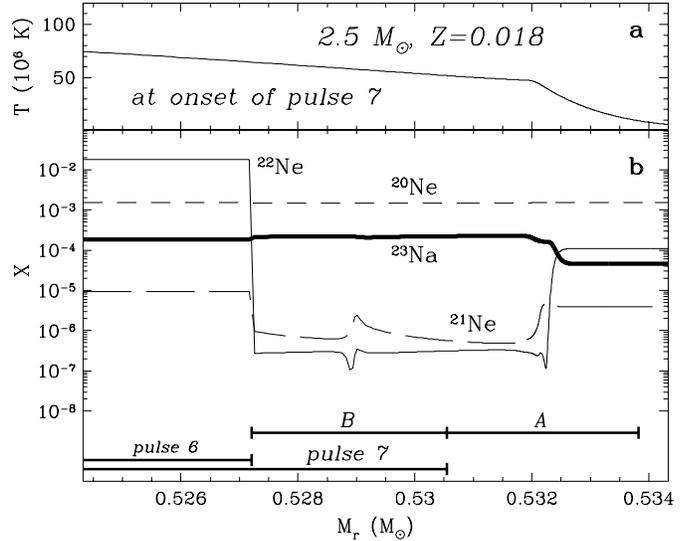}}
  \caption{{\bf a} Temperature and {\bf b} abundance profiles
           of the stable Ne-Na nuclei, as labeled on the curves,
           in the intershell layers at the end of the 6th interpulse
           of the \mass{2.5}, $Z=0.018$ standard model star. The extensions of zones
           {\sl A} and {\sl B} described in Sect.~\ref{Sect:AGB} and
           of pulses before and after the considered interpulse are also shown
           in the lower part of panel {\bf b}.
          }
  \label{Fig:Nap7}
\end{figure}

\begin{figure}
  \resizebox{\hsize}{!}{\includegraphics{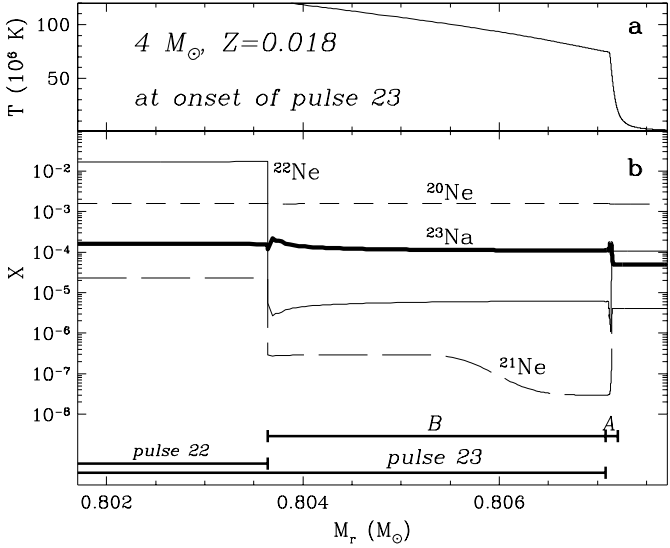}}
  \caption{Same as Fig.~\ref{Fig:Nap7}, but at the end of the 22nd interpulse
           of the \mass{4}, $Z=0.018$ standard model star.
          }
  \label{Fig:Nap23}
\end{figure}

\subsection{Intershell $^{23}Na$ abundances during the AGB phase}
\label{Sect:Na23standard}

As recalled in the preceding section,
no surface \chem{Na}{23} abundance alteration is predicted by our standard
TP-AGB models\footnote{The eventual
dredge-up of the He-rich (but not C-rich) material from zone $A$ (Fig.~\ref{Fig:phases})
leads to a negligible alteration of the surface Na abundance.}.
It is, however, instructive to analyze the abundance
profile of the {\it secondary} \chem{Na}{23} in the intershell layers of the standard
models. The question of \chem{Na}{23} destruction by p-capture, in particular, is of
relevance for our study.

The intershell temperature and Ne-Na abundance profiles at the end of the 6th
interpulse of the \mass{2.5}, $Z=0.018$ model star are shown in Fig.~\ref{Fig:Nap7}.
Secondary \chem{Na}{23} is produced at the level expected from the operation of the
NeNa chain (Sect.~\ref{Sect:Na23scenario}). The temperatures characterizing the HBS
are lower than $50\times 10^6$~K, and no \chem{Na}{23} destruction by p-capture is
obtained, as expected. The intershell \chem{Na}{23} abundance is 5 times higher
than its surface abundance.

The same profiles are presented in Fig.~\ref{Fig:Nap23}, but for the 22nd
interpulse of the \mass{4}, $Z=0.018$ model star. The HBS of those models is
characterized by temperatures reaching $75\times 10^6$~K, leading to the activation
of \chem{Na}{23} burning by p-capture. Indeed, a slight destruction of
\chem{Na}{23} is observed in the ashes left over by the HBS (Fig.~\ref{Fig:Nap23}).
However, the intershell \chem{Na}{23} abundance is found to still
be 2.3 times higher than its
surface abundance. Compared to the intershell abundance in the 6th interpulse of
the \mass{2.5}, $Z=0.018$ star, the \chem{Na}{23} destruction amounts to
a factor of about 2.
This Na destruction factor agrees with the yields of the parametric H-burning
calculations presented in Appendix~A.

Recently, a new compilation of nuclear reaction rates has been provided by
the NACRE project (Arnould et al.~\cite{Arnould_etal99}), which
predicts much higher \chem{Na}{23} p-capture rates than estimated
before. In order to evaluate the impact of those new reaction rates on
Na production in AGB stars, the 22nd interpulse of the \mass{4}, $Z=0.018$ is
recomputed with those new rates. The resulting intershell \chem{Na}{23}
abundance amounts now to only 70\% of its surface abundance. The destruction
factor is thus 7, which is much higher than with the old reaction rates.
It must be noted, however, that the high uncertainties still
affecting the p-capture rates do not, at present time, enable to
lead to a definite conclusion. Arnould et al.~(\cite{Arnould_etal99})
show, for example, that \chem{Na}{23} can still be produced to an abundance
3 times higher than its initial abundance, even at $80\times 10^6$~K, within the
uncertainties of the new nuclear reaction rates.

\subsection{Intershell \chem{C}{12} and \chem{Ne}{22} abundances}
\label{Sect:Ne22}

\begin{figure}
  \resizebox{\hsize}{!}{\includegraphics{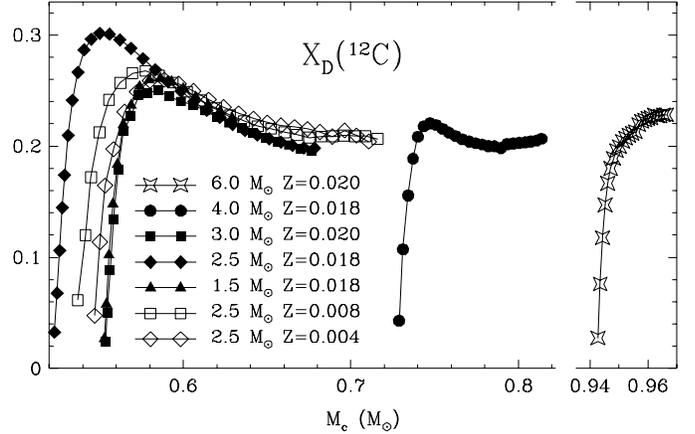}}
  \caption{\chem{C}{12} mass fraction
           emerging from the pulses of our model stars
           as a function of the H-depleted core mass.
           Each mark locates a pulse.
          }
  \label{Fig:XpC12}
\end{figure}

\begin{figure}
  \resizebox{\hsize}{!}{\includegraphics{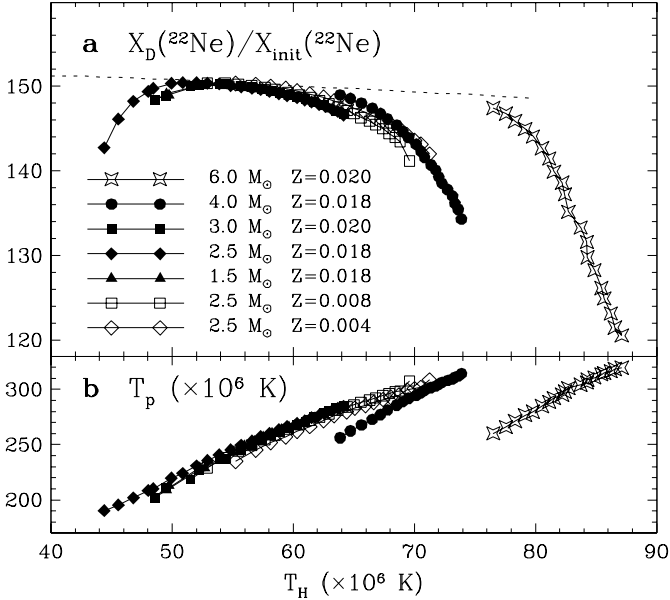}}
  \caption{{\bf a} Abundance of \chem{Ne}{22}, relative to its initial stellar
           abundance, emerging {\it from the fifth pulse on} in our model stars
           as a function of the temperature of the H-burning shell
           (taken at the layer of maximum energy production
           and at the onset of each pulse).
           Each mark locates a pulse.
           {\bf b} Same as {\bf a}, but for the maximum temperature
           reached at the base of the pulses.
          }
  \label{Fig:XpNe22}
\end{figure}

The mass fraction of \chem{C}{12} emerging from the pulses is shown in
Fig.~\ref{Fig:XpC12} for all our model stars as a function of the H-depleted
core mass. It reaches up to 30\% in the \mass{2.5}, $Z=0.018$ star, and evolves
after a sufficient number of pulses to an asymptotic value of $\sim 20$\%
independent of the stellar mass and metallicity.
The abundance of \chem{Ne}{22} emerging from the pulses, on the other hand,
is shown in Fig.~\ref{Fig:XpNe22}a as a function of the temperature of
the HBS at the onset of each pulse.

Let us consider the \chem{Ne}{22} production in the intershell layers.
If we assume that all the available \chem{C}{12} and \chem{O}{16} in the
HBS are transformed into \chem{N}{14}, then $1.24\times 10^{-2}$ mass
fractions of \chem{N}{14} are expected in the intershell layers for
solar initial chemical composition (cf. Table~\ref{Tab:initial abundances}).
Converted into \chem{Ne}{22}, this would
lead to a \chem{Ne}{22} abundance emerging from each pulse,
$X_D(\mchem{Ne}{22})$, 157 times higher than its initial abundance
$X_{init}(\mchem{Ne}{22})$. For solar initial composition, this corresponds to
$X_D(\mchem{Ne}{22})=0.019$ mass fractions.

In reality, \chem{C}{12} and
\chem{O}{16} are present at the level of few percents in the ashes of
H-burning, reducing accordingly  the amount of \chem{N}{14} available
for the synthesis of \chem{Ne}{22}. Though the reduction is very small
(few percents), it is instructive to pursue further the discussion.
Parametric nucleosynthesis calculations, presented in Appendix~A,
show that the abundance of \chem{N}{14} emerging from H-burning actually
depends slightly on the temperature at which hydrogen burns.
Taking those temperature-dependent \chem{N}{14} yields, the expected
$X_D(\mchem{Ne}{22})/X_{init}(\mchem{Ne}{22})$ ratio ranges between 149 and
151 (dotted line in Fig.~\ref{Fig:XpNe22}a). Comparing that value with the
intershell \chem{Ne}{22} abundance predicted by the stellar models (filled and open symbols in
Fig.~\ref{Fig:XpNe22}), we see that essentially all \chem{N}{14}
transforms into \chem{Ne}{22} at pulse temperatures $T_p$ (given
in Fig.~\ref{Fig:XpNe22}b) exceeding $\sim 210\times 10^6$~K. Such temperatures
are already reached in the very first pulses of most of our model
stars. The assumption of a complete conversion of \chem{N}{14} into
\chem{Ne}{22} in the intershell layers is thus very good.

At $T_p$ exceeding $250-280\times 10^6$~K, on the other hand, Fig.~\ref{Fig:XpNe22}
shows that the \chem{Ne}{22} abundances predicted by the standard models
are lower than that expected from the total conversion of the \chem{N}{14}
left over by the HBS. This is explained by the partial destruction of
\chem{Ne}{22}, which begins to burn by $\alpha$ capture.
The \chem{Ne}{22} destruction factor, however, is less than 5\% in all our stars,
except in the 4 and \mass{6} stars where it respectively reaches 10 and 20\%
in the 24th pulse (with $T_p=312$ and $320\times 10^6$~K, respectively). Such a destruction
does not have significant implication on the \chem{Na}{23} predictions for the
\mass{4} star, though, and is neglected in the synthetic and analytical calculations
presented in Sect.~\ref{Sect:Na23}. The case of the \mass{6} star, on the other
hand, is discussed in Sect.~\ref{Sect:HBB}.

\section{Sodium abundance predictions in models with dredge-up}
\label{Sect:Na23}

In this section, the production of sodium is analyzed in those AGB stars which do not
experience HBB. The case of stars experiencing HBB is considered in
Sect.~\ref{Sect:HBB}.

\subsection{Synthetic calculations: the method}
\label{Sect:synthetic calculations}

\begin{table*}
\caption[]{\label{Tab:Xat1stpulse}
           Surface mass fractions of the \chem{C}{12}, \chem{N}{14}, \chem{O}{16},
           \chem{Ne}{22} and \chem{Na}{23} predicted by the standard model stars at the
           onset of the TP-AGB phase. The total C+N+O mass fraction is also given in the 4th
           line.  The increase in those mass fractions
           relative to their initial values, due to the operation of the first and,
           possibly, second dredge-ups, are given in parentheses (in dex).
          }
\begin{tabular}{cccccccc}
\hline
\noalign{\smallskip}
  &  \mass{4} &  \mass{3} & \mass{2.5} & \mass{1.5} & \mass{2.5} & \mass{2.5}\\
  & $Z$=0.018 & $Z$=0.020 &  $Z$=0.018 &  $Z$=0.018 &  $Z$=0.008 &  $Z$=0.004\\
\noalign{\smallskip}
\hline
\noalign{\smallskip}
 \chem{C}{12}  & 1.785e-3 ($-$0.21) & 2.013e-3 ($-$0.20) & 1.820e-3 ($-$0.20) & 2.052e-3 ($-$0.15) & 7.714e-4 ($-$0.22) & 3.660e-4 ($-$0.24)\\
 \chem{N}{14}  & 2.661e-3 ($+$0.40) & 2.823e-3 ($+$0.38) & 2.421e-3 ($+$0.36) & 1.947e-3 ($+$0.27) & 1.166e-3 ($+$0.40) & 6.181e-4 ($+$0.42)\\
 \chem{O}{16}  & 8.632e-3 ($-$0.02) & 9.699e-3 ($-$0.02) & 8.849e-3 ($-$0.01) & 9.100e-3 ($-$0.00) & 3.882e-3 ($-$0.02) & 1.926e-3 ($-$0.02)\\
  CNO          & 1.308e-2 ($+$0.00) & 1.454e-2 ($+$0.00) & 1.309e-2 ($+$0.00) & 1.310e-2 ($+$0.00) & 5.819e-3 ($+$0.00) & 2.910e-3 ($+$0.00)\\
\noalign{\medskip}
 \chem{Ne}{22} & 1.071e-4 ($-$0.06) & 1.212e-4 ($-$0.05) & 1.105e-4 ($-$0.05) & 1.216e-4 ($-$0.01) & 4.760e-5 ($-$0.06) & 2.311e-5 ($-$0.08)\\
 \chem{Na}{23} & 4.883e-5 ($+$0.19) & 5.200e-5 ($+$0.17) & 4.536e-5 ($+$0.16) & 3.370e-5 ($+$0.03) & 2.171e-5 ($+$0.19) & 1.156e-5 ($+$0.22)\\
\noalign{\smallskip}
\hline
\end{tabular}
\end{table*}

  In order to predict the surface abundance evolution of \chem{C}{12},
\chem{Ne}{22} and \chem{Na}{23} in our model stars, we have to use `synthetic'
calculations since standard models cannot reproduce consistently the
3DUP phenomenon (Sect.~\ref{Sect:models}). In such synthetic calculations,
the chemical evolution of a star is computed by estimating {\it a posteriori}
the effects of successive 3DUP episodes on the intershell and surface
abundances. In doing so, we assume that the overall structural evolution of the model
stars is not affected by the operation of those 3DUPs. The evolution, as a
function of pulse number, of the mass $M_{env}$ of the envelope (Fig.~10
of MM99), of the masses $M_A$, $M_B$ and $M_D$ of zones {\it A}, {\it B} and {\it D},
respectively (Fig.~8 of MM99), and of the mass fraction $^{12}\!X^i_D$
of \chem{C}{12} emerging from each pulse (Fig.~\ref{Fig:XpC12}),
are thus taken from the standard stellar model predictions.
The chemical evolution of the intershell layers and of the envelope are then
computed by considering their abundance changes during each
interpulse/pulse/afterpulse phases.

  Let us consider pulse $i$. The abundances of
\chem{N}{14} and \chem{Na}{23} available in zone {\it B} ($^{14}X^i_B$ and
$^{23}X^i_B$, respectively) are determined by the nucleosynthesis occurring in
the HBS during the preceding interpulse $i-1$. We assume that all available
\chem{C}{12} and \chem{O}{16} are converted in the HBS into \chem{N}{14} through the
CNO cycles, and that all available \chem{Ne}{22} is burned into \chem{Na}{23}
by the NeNa chain.
The destruction of \chem{Na}{23} by p-captures (at $T>60\times 10^6$~K) is taken into
account by a destruction factor
$d \equiv X(\mchem{Na}{23})/[X_0(\mchem{Na}{23})+\frac{23}{22}X_0(\mchem{Ne}{22})]$
as defined in Appendix~A. Its value is determined in that
Appendix as a function of temperature at which H burns, taken here to be the
temperature of the HBS predicted by the standard model predictions (Fig.~\ref{Fig:XpNe22}).
The slight contribution from \chem{Ne}{20} burning, on
the other hand, is neglected since \chem{Ne}{20} is not produced in the intershell
layers. This assumption imposes a maximum value of $d=1$.
We then have
\begin{equation}
\label{Eq:14XB}
  {^{14}\!X^i_B} = \frac{14}{12}\times {^{12}\!X^{i-1}_s}
                   + {^{14}\!X^{i-1}_s}
                   + \frac{14}{16}\times {^{16}\!X^{i-1}_s}
\end{equation}
\label{Eq:23XB}
\begin{equation}
  {^{23}\!X^i_B} = d\; \left( \frac{23}{22}\times {^{22}\!X^{i-1}_s}
                              + {^{23}\!X^{i-1}_s}
                       \right)
\end{equation}
where $^{12}\!X^{i-1}_s$, $^{22}\!X^{i-1}_s$ and $^{23}\!X^{i-1}_s$
are respectively the abundances of \chem{C}{12}, \chem{Ne}{22} and \chem{Na}{23}
in the envelope during interpulse $i-1$.

During pulse $i$, we assume that all \chem{N}{14} available in zone {\it B} is
transformed into \chem{Ne}{22}. In doing so, we neglect the
fact that the transformation is incomplete in the coolest first pulses of some
model stars. We further assume that the amount of \chem{Ne}{22} burned in the
hottest pulses of our stars is negligible (see Sect.~\ref{Sect:Ne22}).
Knowing that sodium remains unaffected by He-burning in the pulse, the abundances of
\chem{Ne}{22} and \chem{Na}{23} emerging in zone {\it D} are respectively given
by
\begin{equation}
\label{Eq:22XD}
  {^{22}\!X^i_D} = \frac{22}{14}\times {^{14}\!X^i_B} \times \frac{M^i_B}{M^i_D}
\end{equation}
\begin{equation}
\label{Eq:23XD}
  {^{23}\!X^i_D} = \frac{ {^{23}\!X^i_B} \times M^i_B + {^{23}\!X^{i-1}_D} \times (M^i_D-M^i_B) }
                        {M^i_D}
\end{equation}

The 3DUP process during afterpulse $i$ then mixes part of the material of zone
{\it D} into the envelope. The mass $M_{dup}$ of material dredged-up from zone
{\it D} is usually expressed relative to the core mass increase $\Delta M_c$
during the preceding interpulse by the `dredge-up efficiency' parameter
$\lambda \equiv M_{dup}/\Delta
M_c$. A value of $\lambda=1$, for example, means that the amount of mass
dredged-up is equal to the amount of H-rich material burned during the interpulse
(the core mass would then remain constant from one pulse to the next). Since $\Delta
M_c=M_B$, the surface abundance of, for example, \chem{Na}{23} after dredge-up is
given by
\begin{equation}
\label{Eq:12Xenv}
  {^{23}\!X^i_s} = \frac{{^{23}\!X^{i-1}_s} \times M^i_{env} +
                         {^{23}\!X^i_D} \times \lambda \times M^i_B}
                         {{M^i_{env}} + \lambda \times M^i_B}.
\end{equation}
Similar relations hold for the surface abundances of \chem{C}{12} and \chem{Ne}{22}.

  Dredge-up is assumed to operate from the 5th pulse on in all the stars,
and Eqs.~(\ref{Eq:14XB}) to (\ref{Eq:12Xenv}) are solved pulse after pulse for
the abundances in zones {\it B}, {\it D} and in the envelope. $M_{env}$,
$M_B$, $M_D$ and $^{12}\!X_D$ are taken from the standard AGB model predictions.
We still need to
specify the surface abundances at the onset of the thermally pulsing AGB (TP-AGB)
phase and the dredge-up efficiency. The abundances at the onset of the TP-AGB
phase are taken from the standard model predictions as summarized in
Table~\ref{Tab:Xat1stpulse}. The dredge-up efficiency, on the other hand,
is a function of stellar mass, metallicity and pulse number. Its value however
is still a matter of research.
Fortunately, the exact value of $\lambda$ is not crucial for our purpose,
as long as the surface \chem{Na}{23} abundance is expressed as a function of the
surface \chem{C}{12} (or C+N+O) abundance (see Sect.~\ref{Sect:Na23discussion}).
For simplicity, we thus take $\lambda=1$.

\subsection{Sodium abundance predictions}
\label{Sect:Na23predictions}

\begin{figure}
  \resizebox{\hsize}{!}{\includegraphics{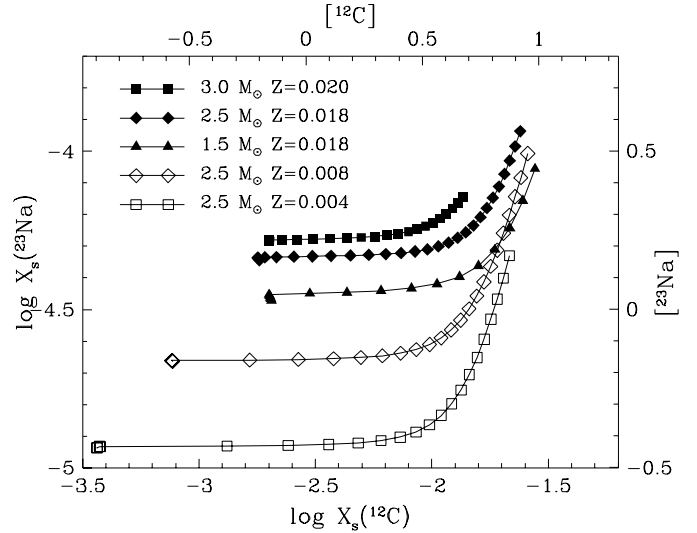}}
  \caption{Surface \chem{Na}{23} versus \chem{C}{12} mass fraction (in logarithm)
           predicted by synthetic calculations (Sect.~\ref{Sect:Na23predictions}),
           for the models as labeled in
           the figure. The abundances expressed in dex are also shown in
           the upper and right axis, with [X]$\equiv$log(X/X$_\odot$).
           The \mass{4}, $Z=0.018$ synthetic star shows little variation in
           its surface abundances (at most 0.3~dex variation in [\chem{C}{12}]
           between the first and 24th pulse with the parameters used in the
           synthetic calculations), and is not shown in the figure for clarity.
          }
  \label{Fig:XsCNa}
\end{figure}

  The surface \chem{Na}{23} abundance evolution predicted by the synthetic
calculations outlined in Sect.~\ref{Sect:synthetic calculations}
are shown in Fig.~\ref{Fig:XsCNa} as a function of the surface \chem{C}{12}
abundance (each mark in the figure locates a pulse). The carbon abundance is
seen to increase steadily with pulse number, as expected from the operation of
the 3DUP episodes. The sodium abundance, on the other hand, remains unaffected during
the first pulses, but steadily increases at more advanced pulses.

We can roughly estimate the increase in the \chem{Na}{23} surface abundance
($\Delta {^{23}\!X_s}$) expected to result from given CNO abundances
in the envelope. As explained in
Sect.~\ref{Sect:Na23scenario}, the process of surface \chem{Na}{23} enhancement from
the envelope's CNO abundances requires two consecutive 3DUPs, one to
dredge-up \chem{Ne}{22} synthesized in the H+HeBS from the CNO elements,
and one to dredge-up \chem{Na}{23} synthesized from \chem{Ne}{22} during the
following interpulse.
At first guess, the resulting increase in the surface sodium abundance would thus be
\begin{equation}
\label{Eq:XenvNa23,dup}
  \Delta {^{23}\!X_s} \approx  d \times f^2
                                 \times \frac{23}{14} \times {^{C\!N\!O}Y_s},
\end{equation}
where $f\equiv M_{dup}/M_{env}$ is the dilution factor suffered by
the C-rich material dredged-up in the envelope, $d$ the \chem{Na}{23}
destruction factor (Sect.~\ref{Sect:synthetic calculations}), and
\begin{equation}
\label{Eq:ys}
  {^{C\!N\!O}Y_s} \equiv \frac{14}{12}\;{^{12}\!X_s} + {^{14}\!X_s} + \frac{14}{16}\;{^{16}\!X_s}
\end{equation}
is the equivalent \chem{N}{14} mass fraction of the CNO abundances in
the envelope (i.e. the \chem{N}{14} mass fraction which would result from the total
transformation of the CNO elements into \chem{N}{14}).
${^{12}\!X_s}$, ${^{14}\!X_s}$ and ${^{16}\!X_s}$
are respectively the surface mass fractions of \chem{C}{12}, \chem{N}{14} and
\chem{O}{16}.
Taking, for example, $d=1$, $f=0.005$ and solar CNO abundances, one gets
$\Delta {^{23}\!X_s}\simeq 5\times 10^{-7}$. This is 60 times lower than the solar
sodium abundance, so that no significant enhancement in the surface
\chem{Na}{23} abundance would be predicted. This conclusion holds true at all
metallicities
if we assume an initial solar distribution of the elements heavier than
helium.

  The process of \chem{Na}{23} production in the envelope is, however,
non linear, and Eq.~(\ref{Eq:XenvNa23,dup}) becomes {\it invalid} after the
operation of few 3DUPs.
This results from the fact that \chem{Ne}{22} builds up in the envelope dredge-up
after dredge-up.
Since \chem{Ne}{22} is the seed nucleus for the production of \chem{Na}{23},
the surface \chem{Na}{23} abundance ${^{23}\!X_s}$ would depend quadratically
on the surface \chem{Ne}{22} abundance ${^{22}\!X_s}$
(a constant ${^{22}\!X_s}$ would lead to a
linear increase of ${^{23}\!X_s}$ with pulse number). If we further
take into account the increase in the envelope's carbon mass fraction pulse after pulse
due to the dredge-up of primary \chem{C}{12},
then ${^{23}\!X_s}$ becomes roughly proportional to $({^{12}\!X_s})^3$
(see Appendix~B). This
explains the rapid increase of the surface \chem{Na}{23} abundance as the
surface carbon abundance increases.

Figure~\ref{Fig:XsCNa} further reveals that the surface sodium abundance
enhancements are higher in stars with lower metallicities. This is due to the
primary nature of \chem{Na}{23} dredged-up to the surface. Actually, the
[\chem{C}{12}]$-$[\chem{Na}{23}] predictions become independent of metallicity
after the operation of a sufficient number of 3DUPs. For example,
the surface mass fraction of \chem{Na}{23} reaches $\sim 10^{-4}$ after
about 25 3DUPs in the \mass{2.5} stars {\it regardless} of their metallicity
(Fig.~\ref{Fig:XsCNa}).

\subsection{Discussion}
\label{Sect:Na23discussion}

\begin{figure}
  \resizebox{\hsize}{!}{\includegraphics{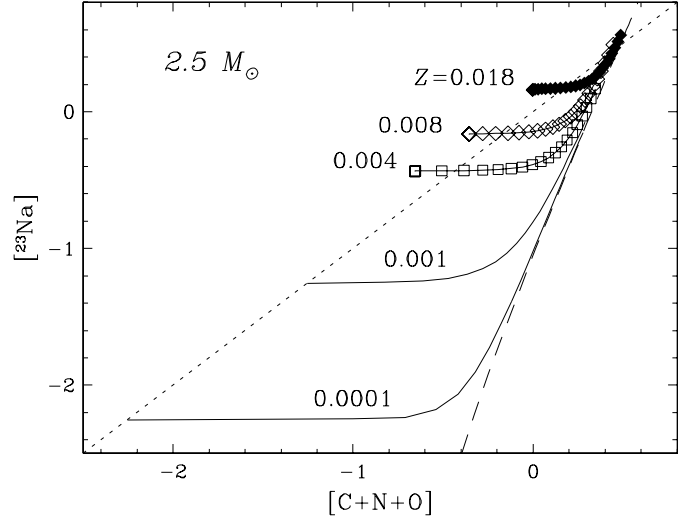}}
  \caption{Same as Fig.\ref{Fig:XsCNa}, but for a \mass{2.5}
           star at different metallicities. The predictions at
           $Z=0.018$, 0.008 and 0.004 are from synthetic
           calculations (Sect.~\ref{Sect:Na23predictions}), while
           those at $Z=0.001$ and 0.0001 are from analytical
           evaluations (Sect.~\ref{Sect:Na23discussion}).
          }
  \label{Fig:XsCNONaZ}
\end{figure}

 It is possible to further study the production of \chem{Na}{23} in AGB stars
{\it analytically} if we assume the constancy of the following four parameters:
the intershell \chem{C}{12} mass
fraction $^{12}\!X_D$ dredged-up to the surface during each 3DUP episode, the
dilution fraction $f$ suffered by the C-rich material when dredged-up from the
intershell layers into the envelope, the dilution factor $h\equiv M_B/M_D$
suffered by the ashes of HBS (\chem{N}{14} in particular) when mixed from zone $B$
of the intershell layers into the pulse, and the destruction factor $d$ of
\chem{Na}{23} by p-captures. The analytical derivation
is performed in Appendix~B. It enables, in particular,
to disentangle the effects of each of the four parameters,
to analyze the influence of metallicity (Sect.~\ref{Sect:Na23Z}) and stellar mass
(Sect.~\ref{Sect:Na23M}) on the sodium predictions,
and to estimate the uncertainties affecting those predictions
(Sect.~\ref{Sect:Na23uncertainties}).

The two main conclusions of the analytical study performed in Appendix~B are:

\vskip 1mm
\noindent  a) primary \chem{Na}{23} production in {\it zero} metallicity stars\footnote{
The evolution of $Z=0$ stars might qualitatively be quite different than that of
$Z\ne 0$ stars. However, the only requirement for our sodium production scenario to
work is the occurrence of 3DUPs which bring primary carbon into the envelope. And
model predictions confirm the existence of thermal instabilities in the He-burning
shell of AGB stars -- and thus most probably of 3DUP events -- at metallicities
down to $Z$=0 for $M\simlr\mmass{4}$ (Fujimoto et al. \cite{Fujimoto_etal84}).
In any case, the LOPSE could very well be defined by the
asymptotic evolution of $^{23}\!X_s$ in a very low (non zero) metallicity
star.
\label{NOTE:Z=0}
}
follows a specific track in the [C+N+O]$-$[\chem{Na}{23}] diagram (with
[C+N+O]=[$^{C\!N\!O}Y_s$]), given by
Eq.~(\ref{Eq:c-s,Z=0}) in the Appendix and represented by a long-dashed line in
Fig.~\ref{Fig:XsCNONaZ}. It is called the {\it line of primary sodium enrichment}
(LOPSE), and is, to first order, a straight line of slope 3
[cf. Eq.~(\ref{Eq:c-s,Z=0,f=0}) in the Appendix];

\vskip 1mm
\noindent b) in {\it non-zero} metallicity stars, the surface \chem{Na}{23}
abundance is given by Eq.~(\ref{Eq:c-s}) in the Appendix. $^{23}\!X_s$ is predicted to
keep constant during the first 3DUPs until the abundance of primary \chem{Na}{23} produced
by the star (and given by the LOPSE) reaches a level comparable to the initial surface
\chem{Na}{23} abundance. It then increases steadily, and approaches the LOPSE asymptotically.

\vskip 1mm
  The evolutionary tracks predicted at $Z=10^{-4}$ and $10^{-3}$ in the
[C+N+O]$-$[\chem{Na}{23}] diagram by Eq.~(\ref{Eq:c-s}) of the Appendix, for
example, are shown by the two unmarked solid lines in
Fig.~\ref{Fig:XsCNONaZ}. Those analytical predictions are in good agreement
with the $^{23}\!X_s$ evolution predicted by the synthetic calculations.
For comparison, the synthetic predictions for the \mass{2.5} stars at
$Z=0.004$, 0.008 and 0.018 are also drawn in Fig.~\ref{Fig:XsCNONaZ}
(symbol-marked solid lines);

\vskip 1mm
  Appendix~B further analyses the sensitivity of the analytical \chem{Na}{23}
abundance predictions to $^{12}\!X_D$, $f$, $h$ and $d$.
The value of $^{12}\!X_D$ predicted by
our standard models has been analyzed in Sect.~\ref{Sect:Ne22}. It reaches
a maximum value of 25-30\% (depending on the initial stellar
mass) after a dozen of pulses, and asymptotically decreases thereafter to
a `canonical' value of $^{12}\!X_D=0.20$ (Fig.~\ref{Fig:XpC12}). This value
is the one adopted in the curves at $Z=10^{-4}$ and $10^{-3}$ displayed in
Fig.~\ref{Fig:XsCNONaZ}. The analysis of Appendix~B reveals a strong sensitivity
of the LOPSE on $^{12}\!X_D$, though. This results from a dependence of $^{23}\!X_s$ on
the square of $^{12}\!X_D$ [cf. Eq.~(\ref{Eq:c-s,Z=0,f=0}) in the Appendix].
Increasing $^{12}\!X_D$ by a factor of two, from
0.2 to 0.4 for example, decreases $^{23}\!X_s$ predicted by the LOPSE
at a given surface C+N+O abundance by a factor of about four (0.6~dex;
Fig.~\ref{Fig:analXfhd}a of Appendix~B). Fortunately, the value of $^{12}\!X_D$
after a dozen of pulses is rather well constraint according to standard stellar
models, lying between 0.20 and 0.24. Furthermore, the value of
$^{12}\!X_D$ is rather independent of metallicity (Sect.~\ref{Sect:Ne22}).
Assuming a constant value for that parameter is thus a good enough
assumption for our purposes. Future evolutionary AGB model calculations
consistently reproducing the 3DUP processes should lead to more refined
predictions.

\begin{figure}
  \resizebox{\hsize}{!}{\includegraphics{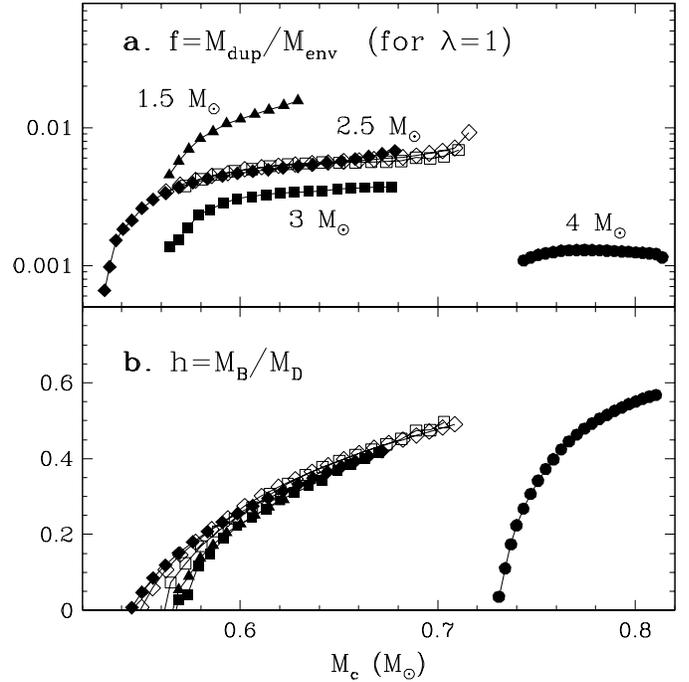}}
  \caption{{\bf a} Dilution factor $f=M_{dup}/M_{env}$, as a function of core mass,
           suffered by the material dredged-up by a 3DUP episode from the C-rich
           intershell layers into the envelope,
           as predicted by the standard AGB
           models when a dredge-up efficiency of $\lambda=1$ is assumed. The
           symbols identifying the mass and metallicity of each line are the same
           as those Fig.~\ref{Fig:XpC12}.
           {\bf b} Same as {\bf a}, but for the dilution factor $h=M_B/M_D$
           suffered by the ashes of the hydrogen burning shell when mixed into
           the pulses.
          }
  \label{Fig:fh}
\end{figure}

  The second parameter is the dredge-up dilution factor $f$. Appendix~B reveals that
$^{23}\!X_s$ predictions are not very sensitive to $f$.
To first order, the LOPSE is even independent on that parameter
[Eq.~(\ref{Eq:c-s,Z=0,f=0}) in the Appendix]. This
independence of $^{12}\!X_s$ on $f$ is welcomed since the dredge-up efficiency
as a function of mass and metallicity is still a matter of research. We choose
a `canonical' value $f=0.005$. This value corresponds to that predicted by
the standard \mass{2.5} models if we assume a dredge-up efficiency of
$\lambda=1$, as shown in Fig.~\ref{Fig:fh}a.

  The third parameter is dilution factor $h$ of H-burning ashes into pulses.
Standard AGB models predict a continuous increase
of that parameter along the AGB, from 0.05 to 0.50-0.60 as reported in
Fig.~\ref{Fig:fh}b. We use a `canonical' value of $h=0.40$.
The analysis developed in Appendix~B, however, reveals a linear dependence of
$^{23}\!X_s$ to $h$ on the LOPSE [Eq.~(\ref{Eq:c-s,Z=0,f=0}) in the Appendix].
As a result, an increase of $h$ from 0.2 to 0.4, for example, would result in an
increase of $^{23}\!X_s$ by a factor of two at a given $^{12}\!X_s$.
Giving a realistic range 0.20-0.60 for $h$ in AGB models, the error on
$^{23}\!X_s$ due to $h$ thus amounts to a factor of three.

  The last parameter is the \chem{Na}{23} destruction factor $d$. With
the nuclear reaction rates used in the
standard model calculations, $d$ drops below 1 only for $T_H>70\times 10^6$~K (cf.
Appendix~A). Because $T_H$ is predicted to be lower than that value in most of our
standard model stars (Fig.~\ref{Fig:XpNe22}), we
adopt a `canonical' value of 1. Since $^{23}\!X_s$ depends linearly on
$d$, the error introduced by using $d=1$ is of a factor of three (see,
however, Sect.~\ref{Sect:Na23uncertainties}).

  The analytical equation (\ref{Eq:c-s}) of the Appendix thus offers a powerful
tool to analyze the production of primary \chem{Na}{23} in AGB stars.
The main {\it qualitative} difference between [C+N+O]$-$[\chem{Na}{23}] predictions
from the analytically-derived relation (\ref{Eq:c-s} in Appendix~B and those
from the synthetic calculations presented in Sect.~\ref{Sect:Na23predictions} (and
which take into account the full stellar characteristics as predicted by standard
AGB models pulse after pulse) is the initial \chem{Na}{23} abundance at the onset
of the TP-AGB phase (see Fig.~\ref{Fig:XsCNONaZ}).
The analytical calculations begin with an initial solar \chem{Na}{23} abundance
scaled to the metallicity (dotted line in Fig.~\ref{Fig:XsCNONaZ}),
while the synthetic calculations take into account
the \chem{Na}{23} abundance enhancements resulting from first and second
dredge-ups.  The increase in $^{23}\!X_s$ due to first dredge-up is, however,
less than 0.3~dex in the stars of interest (cf. Table~\ref{Tab:12DUP}). Its
contribution becomes meaningless when the star reaches the LOPSE, i.e. when primary
\chem{Na}{23} dominates sodium production.

\subsection{Sodium production as a function of metallicity}
\label{Sect:Na23Z}

Figure~\ref{Fig:XsCNONaZ} shows that the increase in the surface C+N+O
abundance required to reach the LOPSE is greater in lower metallicity stars.
At solar metallicity, for example, the surface C+N+O abundance must increase by
less than 0.4~dex before it reaches the LOPSE,
while 1.2~dex are required at $Z=10^{-3}$, and 1.9~dex at $Z=10^{-4}$.
Yet, metal-poor stars are expected to be more
efficient in producing primary \chem{Na}{23} than solar-metallicity stars. This results from the
fact that the initial surface C+N+O abundance is proportional
to the metallicity (the CNO elements represent 72\% of the metallicity),
while the primary \chem{C}{12} mass fraction produced in the HeBS is independent
of metallicity (Fig.~\ref{Fig:XpC12}). As a result, the {\it relative} increase
of the surface \chem{C}{12} abundance is higher in metal-poor stars. We can easily
estimate the number $n_{dup}$ of 3DUPs necessary to increase the surface
C+N+O abundance by a factor of, let us say, $x$. Let us assume for simplicity
that the dredge-up dilution factor $f$ is constant from one pulse to the next. We
then have
\begin{equation}
\label{Eq:ndup}
  n_{dup} = (x-1) \; \frac{0.72 \times Z}{{^{12}\!X_D} \times f}.
\end{equation}
For example, the number of 3DUPs necessary to increase $^{C\!N\!O}Y_s$ by a factor of
three (i.e. by 0.5~dex) is, assuming $^{12}\!X_D=0.24$ and $f=0.006$,
\begin{displaymath}
\label{Eq:ndupZ}
  n_{dup} \simeq 1000\;Z.
\end{displaymath}
This corresponds to 18 dredge-ups in a solar metallicity star, 8 at $Z=0.008$,
and only one at $Z\leq 0.001$. The efficiency of surface C enrichment
thus increases very rapidly with decreasing metallicity.
{\it Low-metallicity stars should thus be good candidates for the production of primary
$^{23}\!Na$}.

  At $Z=10^{-4}$, for example, the number of 3DUPs required for
$^{C\!N\!O}Y_s$ to reach the LOPSE
(i.e. to increase by 1.9~dex) is expected to be 4 according to
Eq.~(\ref{Eq:ndup}). Of course, Eq.~(\ref{Eq:ndup}) only provides an order
of estimate, and those predictions should be
confirmed by future evolutionary AGB models {\it consistently} reproducing the 3DUPs.
In particular, the dependence law of the dredge-up efficiency on mass, metallicity
and pulse number is of crucial importance in determining the
number of pulses required to reach the LOPSE. According to current estimations
the dredge-up efficiency is expected to increase with decreasing metallicity. This
strengthens our conclusion that low-metallicity stars should be efficient
sites for the production of primary \chem{Na}{23}.

\subsection{Sodium production as a function of stellar mass}
\label{Sect:Na23M}

  The synthetic calculations presented in Sect.~\ref{Sect:Na23predictions} reveal
a dependence of the sodium abundances on the stellar mass.
 Figure~\ref{Fig:XsCNa} indeed shows that $^{23}\!X_s$ begins to increase at a
lower C+N+O abundance in more massive stars. In other words, \chem{Na}{23}
production would be favored in more massive AGB stars.

  Why such a trend in the synthetic predictions? The analytical study performed in
Appendix~B again offers the necessary tool to answer that question.
What actually happens is that in a massive AGB star, the low dilution factor $f$
($M_{env}$ being high) leads to a slow increase of the surface \chem{C}{12} abundance with
pulse number. It takes then many 3DUP episodes for the surface C+N+O abundance
to reach the LOPSE. By that time, the $h$ factor
has already increased to a high value according to standard model predictions
(Fig.~\ref{Fig:fh}b), leading to higher \chem{Na}{23} abundances. In a low-mass AGB star, on
the contrary, the high dilution factors $f$ lead to a rapid increase in the
surface \chem{C}{12} abundance with pulse number. As a result, $h$ is still low
when the surface C+N+O abundance reaches the LOPSE, leading to lower 
$^{23}\!X_s$ values at a given C+N+O surface abundance.

  The $^{23}\!X_s$ predictions thus also depend on the time
at which 3DUP becomes efficient on the TP-AGB (thereby determining $h$),
i.e. on the history of the 3DUP
efficiency. In the synthetic calculations presented in
Sect.~\ref{Sect:synthetic calculations}, it is assumed that 3DUP begins to operate at the 5th pulse
irrespective of the stellar mass and metallicity, and that its efficiency is equal
to $\lambda=1$ for all subsequent pulses. In reality, the dredge-up efficiency is
expected to vary with stellar mass, metallicity, pulse number and mass loss rate.
The lack of reliable predictions, at the present time, for the 3DUP efficiency as
a function of pulse number thus introduces an additional source of uncertainty on
the quantitative prediction of $^{23}\!X_s$.

\subsection{Uncertainties on the \chem{Na}{23} abundance predictions}
\label{Sect:Na23uncertainties}

 The quantitative prediction of primary \chem{Na}{23} production in AGB stars has
been shown in Sects.~\ref{Sect:Na23discussion} and \ref{Sect:Na23M} to depend
sensitively on the \chem{C}{12} abundance produced in the HeBS, on the
intershell structural parameter $h$, and on the nuclear destruction factor $d$.

  In our {\it standard} AGB models, the intershell \chem{C}{12} abundance reaches the asymptotic value of 0.20-0.24 after
a sufficient number of pulses (Sect.~\ref{Sect:Ne22}). `Non standard'
input physics, however, may lead to a different conclusion. Herwig et al.
(\cite{Herwig_etal97}), for example, obtain $^{12}\!X_D=0.50$ when applying
extra-mixing (overshooting) at the borders of the convective pulses. Since the
amount of overshooting at the convective pulse borders is
certainly not yet established, we have to consider $^{12}\!X_D$ to be known to only
a factor of 2.5. Moreover, an additional source of discomfort in the quantitative
prediction of $^{12}\!X_D$ results from the uncertainty affecting the nuclear
reaction rate of \reac{C}{12}{\alpha}{\gamma}{O}{16}
(see Arnould et al. \cite{Arnould_etal99}).
The role of $^{12}\!X_D$ is all the more important since the primary \chem{Na}{23} abundance
predicted by the LOPSE depends on the square of $^{12}\!X_D$. If we thus take
a global uncertainty factor of about three on $^{12}\!X_D$ due to convection and
nuclear reaction rates, then the
sodium abundance predicted at a given C+N+O surface abundance is reliable only to
a factor of about 10.

The uncertainties on $h$, on the other hand, have been shown in
Sect.~\ref{Sect:Na23discussion} to lead to an uncertainty factor on $^{23}\!X_s$ of
about 3.

Finally, the uncertainty on $d$ has been shown in
Sect.~\ref{Sect:Na23discussion} to lead to an uncertainty on
$^{23}\!X_s$ of a factor of 3. The new NACRE compilation, however, predict
much higher \chem{Na}{23} destruction rates than before
(Sect.~\ref{Sect:Na23standard}). But the very large uncertainties still affecting
the p-capture rates on \chem{Na}{23} prevent from giving a definite
conclusion about their impact on $^{23}\!X_s$ predictions. A \chem{Na}{23}
destruction factor of `only' 3, for example, is still within the
NACRE uncertainties. We thus adopt this optimistic value, keeping
however in mind a possible higher destruction factor due to nuclear reaction
rates.

  All together, then, the present uncertainty affecting $^{23}\!X_s$ predictions
at a given surface C+N+O abundance amounts to two orders of magnitude.

\subsection{Summary}
\label{Sect:Na23summary}

  Both the predictions of synthetic AGB calculations presented in
Sect.~\ref{Sect:Na23predictions} and the analytic considerations
of Sect.~\ref{Sect:Na23discussion} do support the
scenario described in Sect.~\ref{Sect:Na23scenario} for the production of
primary sodium in AGB stars. The evolution of the surface \chem{Na}{23}
abundance obeys a well defined pattern as a function of the surface C+N+O
abundance, given by Eq.~(\ref{Eq:c-s}) in Appendix~B and shown in
Fig.~\ref{Fig:XsCNONaZ} for several metallicities.

  All AGB stars are thus potential sites for the production of primary
\chem{Na}{23}. For such a production to be efficient, a sufficient
number of 3DUPs must occur in order to increase the surface C+N+O
abundance to the level required by the LOPSE
(dashed line in Fig.~\ref{Fig:XsCNONaZ}) at the given initial $^{23}\!X_s$.
The amount of primary sodium produced in a given
AGB star thus depends on the number of 3DUP episodes experienced by that
star.
Since the dredge-up efficiency increases with decreasing metallicity,
low-metallicity stars are expected to be more efficient in producing
primary \chem{Na}{23}. This conclusion is furthermore strengthened by the fact
that the relative increase in the surface carbon abundance during a 3DUP episode
is higher in low than in high metallicity stars.

 The uncertainties affecting both synthetic and analytic predictions of
$^{23}\!X_s$ are, however, still very large at present time (two orders of
magnitude). In particular, the uncertainties affecting the \chem{Na}{23} burning
reaction rates is of great concern. Eventual observation of \chem{Na}{23} overabundance
in AGB stars and related objects would provide a definite proof of the feasibility
of our scenario for the production of primary \chem{Na}{23}. This question is
briefly addressed in Sect.~\ref{Sect:observations}. Before doing that, however, the
case of massive AGB stars experiencing HBB is discussed in the next section.

\begin{figure}
  \resizebox{\hsize}{!}{\includegraphics{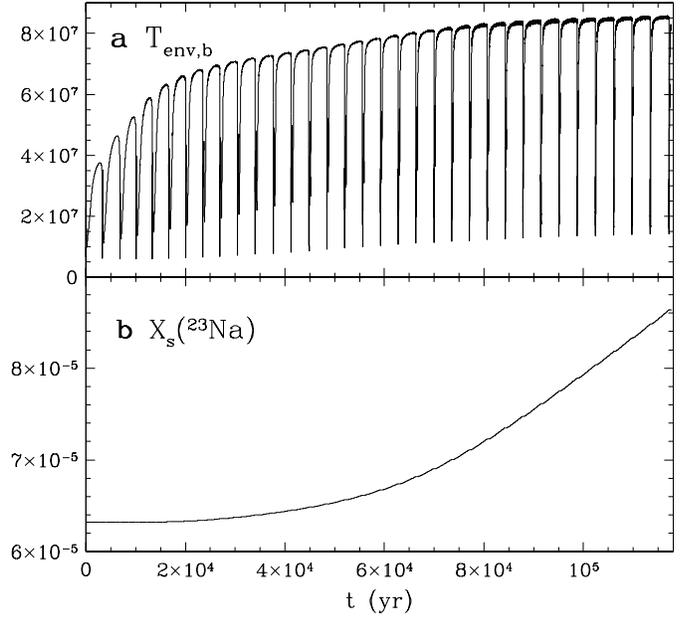}}
  \caption{{\bf a} Evolution of the temperature at the base of the envelope of a
           \mass{6}, $Z=0.018$ standard model star during the first 30 pulses.
           {\bf b} Same as {\bf a}, but for the surface \chem{Na}{23} mass
           fraction. The origin of time is set to the first pulse.
          }
  \label{Fig:Na23hbb}
\end{figure}

\section{Sodium production by hot bottom burning}
\label{Sect:HBB}

  In massive AGB stars ($M\simgr\mmass{5}$ at $Z$=0.02), the temperature at the base
of the envelope may reach high enough values for H-burning to operate {\it in} the
envelope. This process, called HBB, alters the surface abundances of H-burning nuclei
without the need to invoke any dredge-up process. In particular, the activation of the
NeNa chain of H-burning naturally leads to the production of \chem{Na}{23} when the
temperature at the base of the envelope exceeds $60\times 10^6$~K.

\subsection{Secondary sodium}

  In the absence of any 3DUP event, secondary sodium can be synthesized by HBB
from the initial \chem{Ne}{22} present in the envelope.
The evolution of the temperature at the base of the envelope of the \mass{6},
$Z=0.018$ standard model star, for example, and of its surface \chem{Na}{23}
abundance are shown in Fig.~\ref{Fig:Na23hbb} during the first 30 pulses. The
production of secondary sodium is seen to become efficient when the base temperature
increases above $70-80\times 10^6$~K.
The increase in the surface \chem{Na}{23} abundance is steady, eventhough not very
rapid (30\% increase over about 25 pulses).

\subsection{Primary sodium}

In the presence of 3DUPs, the increase in the surface \chem{Na}{23} abundance
through HBB may be considerably higher due to the dredge-up of primary
\chem{Ne}{22}. In such a scenario, only one dredge-up/interpulse/pulse/dredge-up
event is required to dredge-up \chem{C}{12} and process it to \chem{Ne}{22}, the
conversion of \chem{Ne}{22} to \chem{Na}{23} being subsequently performed in the envelope.
The fact that the dredge-up efficiency increases with stellar mass renders this
scenario very attractive. Several important points must however be considered.

First, the standard models reveal a destruction of the intershell \chem{Ne}{22} by
$\alpha$-capture at the He-burning shell temperatures characterizing massive AGB
stars (Sect.~\ref{Sect:Ne22}). In the \mass{6} model star, the destruction factor
amounts to 20\% at the 24th pulse (Fig.~\ref{Fig:XpNe22}). Let us, however, neglect
this destruction factor for now.

Second, the dilution factor $f$ of the \chem{Ne}{22}-rich material mixed from the
intershell layers into the envelope is predicted by standard models to be only 0.0002,
much smaller than in lower mass stars ($f\simeq 0.001$, 0.003, 0.005 and 0.01 in the
4, 3, 2.5 and \mass{1.5} model stars, respectively, see Fig.~\ref{Fig:fh}). As a
result, the surface \chem{Ne}{22} mass fraction is not expected to increase very rapidly
as a result of 3DUP episodes in massive AGB stars. For example, a synthetic
calculation of the surface \chem{Ne}{22} abundance evolution in the \mass{6} model
star, performed with the recipe outlined in Sect.~\ref{Sect:synthetic calculations},
predicts only 20\% increase in the surface \chem{Ne}{22} abundance after 24 pulses.
The synthesis of primary sodium by 3DUP~+~HBB may thus not be very
efficient. A more positive conclusion may however be obtained in low-metallicity stars,
where the low initial \chem{Ne}{22} abundance renders the relative contribution of
the intershell \chem{Ne}{22} to the envelope's \chem{Ne}{22} abundance more important.

Third, synthesizing primary sodium in the intershell layers of massive AGB stars
by two successive dredge-up/interpulse/pulse sequences, in a similar way as in
lower mass stars, is not efficient either.
The temperatures in the H-burning shell of those massive AGB stars
indeed exceed $75\times 10^6$~K according to the predictions from the \mass{6} model star (cf.
Fig.~\ref{Fig:XpNe22}), and lead to the destruction of \chem{Na}{23} by p-capture
in the HBS. At
the end of the 23rd interpulse of the \mass{6} model star, for example, the
\chem{Na}{23} intershell abundance is 30\% lower than its surface abundance.

Finally, we have to consider the effects of mass loss which characterizes AGB stars.
It reduces the mass of the envelope, thereby decreasing the temperature at
its base, and eventually switching HBB off. After how many interpulses is HBB
switched off, however, depends on the still uncertain mass loss rates suffered by
those stars.

The production of primary sodium by HBB seems thus not favored in massive AGB stars.
There are however still too many uncertainties, such as those concerning the dredge-up
efficiency or mass loss rates, to allow a definite conclusion to be drawn.

\section{Observations}
\label{Sect:observations}

The primary sodium, if produced efficiently in AGB stars, should reveal
its imprint in the abundances at the surface of AGB and post-AGB stars.
According to the analysis of Sect.~\ref{Sect:Na23}, sodium could be produced
in AGB stars of nearly solar metallicity when their \chem{C}{12}
abundance has increased by at least 0.7~dex (Fig.~\ref{Fig:XsCNa}). This would correspond to
stars displaying C/O ratios above $\sim$2. Unfortunately, the number of
known such objects is low, due to the fact that those AGB stars are usually
enshrouded into a circumstellar dust which obscures the central AGB
star. To my knowledge, no sodium abundance has so far been measured in
those objects.

The analysis of post-AGB stars seems more promising. Those objects are expected to
display the abundance patterns characterizing AGB stars at the end of their life,
and in particular high C+N+O abundances if 3DUP occurred during
their AGB phase. Spectroscopic analysis of some of those objects has been performed
during the last decade (e.g., Parthasarathy et al. \cite{Parthasarathy_etal92}, Gonzalez
\& Wallerstein \cite{Gonzalez_Wallerstein92}, Za\u{c}s et al. \cite{Zacs_etal96},
Van Winckel et al. \cite{VanWinckel_etal96}, Decin et al. \cite{Decin_etal98}).
In particular, Gonzalez \& Wallerstein (\cite{Gonzalez_Wallerstein92})
analyze the chemical composition of a
post-AGB star, ROA~24, belonging to the globular cluster $\omega$~Cen.
They find significant overabundances of C, N, O, and s-process elements
compared to the abundances measured in other giants in $\omega$~Cen. The fact that
the s-process elements are enhanced is a very strong indication of the post-AGB
nature of that star. Very interestingly, they also find a significant overabundance
of Na in that object relative to the abundance in other giants of $\omega$~Cen. This
would be the first observational evidence of primary \chem{Na}{23} production in
AGB stars. A more thorough analysis of that case and of other post-AGB stars is
outside the scope of this article, though, and is the object of a separate study
(Mowlavi \& Van Winckel, in preparation).

%

\section{Conclusions}
\label{Sect:conclusions}

The main conclusion of this paper is that {\it AGB stars are potential
sites for the production of primary sodium}. For the scenario presented
in Sect.~\ref{Sect:AGB} to be efficient, however, a certain level of
surface C+N+O abundance must be reached by the operation of
sufficient number of 3DUP events (Sect.~\ref{Sect:Na23}).

The surface \chem{Na}{23} abundance evolution can be
predicted as a function of the surface C+N+O abundance by the analytical
relation (\ref{Eq:c-s}) presented in Appendix~B, with `recommended'
values of $^{12}\!X_D=0.20$, $f=1$, $h=0.4$ and $d=1$. The
values of $^{12}\!X_D$, $h$ and $d$ are suggested from standard AGB model
predictions, while the value of $f$ is suggested after the study of Mowlavi
(\cite{Mowlavi99}). The \chem{Na}{23} abundance predictions are
sensitive to $^{12}\!X_D$, $h$ and $d$, but not much on $f$ as long as
$X_s(\mchem{Na}{23})$ is expressed as a function of the C+N+O
abundance.

An important result of the analytical study is the prediction of a
LOPSE in the [C+N+O]$-$[\chem{Na}{23}] diagram,
which provides the \chem{Na}{23} abundance evolution in {\it zero
metallicity} stars as a function of their surface C+N+O abundance. The surface
\chem{Na}{23} abundance in non-zero metallicity stars, on the other hand,
is shown to evolve asymptotically towards the LOPSE.

Hot bottom burning in massive AGB stars may further help the production of \chem{Na}{23}
by synthesizing it directly in the envelope. In that case, only one
dredge-up/interpulse/pulse/dredge-up sequence is necessary to synthesize and
dredge-up primary \chem{Ne}{22}, followed by its burning into \chem{Na}{23} in the
envelope. The analysis presented in Sect.~\ref{Sect:HBB}, however, suggests that
such primary sodium production may not be very efficient in those stars,
at least at solar metallicity. Low metallicity stars
may be better candidates for producing sodium through 3DUP~+~HBB.

The uncertainty on the primary \chem{Na}{23} abundance prediction at a given
C+N+O abundance is presently of two orders of magnitude,
even with synthetic AGB calculations. This poor
prediction partly results from our still poor quantitative knowledge
of the 3DUP efficiency as a function of stellar mass, metallicity and
pulse number, since the values of $^{12}\!X_D$ and $h$ vary with pulse
number in a given star, and partly from the uncertainties affecting the p-capture
rates on \chem{Na}{23}.  Future AGB model predictions consistently
reproducing the 3DUPs, and combined with improved reaction rates, would hopefully
enable to provide a better quantitative prediction of sodium production in low- and
intermediate-mass stars.

Meanwhile, we can search for an observational evidence for this sodium production
scenario in carbon stars and related objects.  Such an evidence may already have been
reported in at least one post-AGB star in the globular cluster $\omega$~Cen
(cf. Sect.~\ref{Sect:observations}).
Further determinations of sodium abundance in those objects and, if possible,
in planetary nebulae are eagerly called for.

  Finally, let me note that the efficient production of primary sodium in
AGB stars, if confirmed by observations, would imply a non negligible
role of those objects in the chemical evolution of galaxies in general, and of
our Galaxy in particular.

\section*{Appendix A: The CNO and Ne-Na modes of hydrogen burning}

\subsection*{The CNO cycles}

\begin{figure}
  \resizebox{\hsize}{!}{\includegraphics{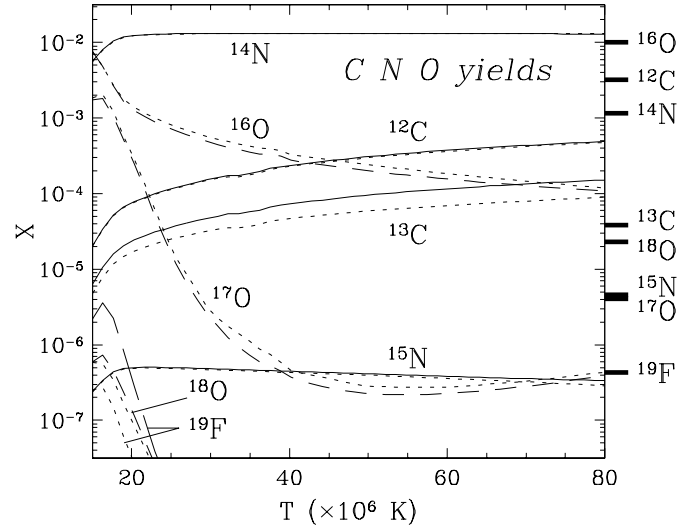}}
  \caption{Mass fractions of the stable nuclides involved in the CNO cycles at
           H exhaustion [$X(\mathrm{H})=10^{-5}$] versus the temperature at which
           hydrogen burns. The dotted lines represent the yields using the NACRE
           reaction rates (see text). The initial (solar) abundances are indicated
           by the horizontal thick lines at the right of the figure.
          }
  \label{Fig:CNOyields}
\end{figure}

The CNO modes of hydrogen burning are well known and documented in the literature
(see, e.g., Arnould et al. \cite{Arnould_etal95}). We only summarize
here their main features in relation with the \chem{Na}{23} production in AGB
stars. The rates of the CNO reactions adopted in the standard model
calculations are the same as in Mowlavi (\cite{Mowlavi95}), except for
\reac{O}{17}{p}{\alpha}{N}{14} and \reac{O}{17}{p}{\gamma}{N}{14} whose rates
are taken from Blackmon et al. (\cite{Blackmon_etal95}).

  The CNO yields obtained at hydrogen exhaustion in one-zone nucleosynthesis
calculations with the temperature assumed constant during the burning are displayed
in Fig.~\ref{Fig:CNOyields} as a function of temperature. As is well known,
essentially all CNO elements are transformed into \chem{N}{14}. However,
\chem{C}{12}, \chem{C}{13} and \chem{O}{16} are still present at H exhaustion at
the level of few percents of the \chem{N}{14} mass fraction.
At temperatures $>40\times 10^6$~K characterizing H-burning in AGB stars, the
mass fraction of \chem{C}{12} (which is the most abundant of the CNO elements
after \chem{N}{14}, see Fig.~\ref{Fig:CNOyields}) increases with temperature.
Concomitantly, the \chem{N}{14} mass fraction at H exhaustion decreases slightly
with temperature (few percents from 40 to over $70\times 10^6$~K). This property is used in
Sect.~\ref{Sect:Ne22} to analyze the \chem{Ne}{22} production in AGB stars.

  The impact of the new NACRE reaction rates (Arnould et al.  \cite{Arnould_etal99})
on the CNO yields of hydrogen burning is shown by the dotted lines in Fig.~\ref{Fig:CNOyields}.
It is seen to be negligible for the \chem{N}{14} production.

\subsection*{The NeNa chain}

\begin{figure}
  \center{
  \resizebox{7.6cm}{!}{\includegraphics{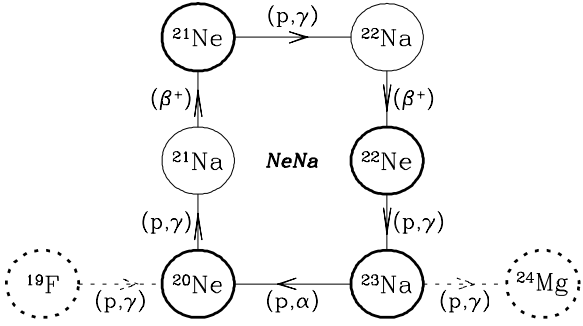}}
  }
  \caption{Reactions of the NeNa chain. Stable nuclides are enclosed in thick
           circles. The possible leakages in and out of the chain are represented by the dashed
           lines.
          }
  \label{Fig:NeNamode}
\end{figure}

  The reactions involved in the NeNa chain of hydrogen burning are shown in
Fig.~\ref{Fig:NeNamode}. The rates of p captures on \chem{Ne}{20}, \chem{Ne}{21},
\chem{Ne}{22} and \chem{Na}{23} are taken from El Eid \&
Champagne (\cite{ElEid_Champagne95}).

\begin{figure}
  \resizebox{\hsize}{!}{\includegraphics{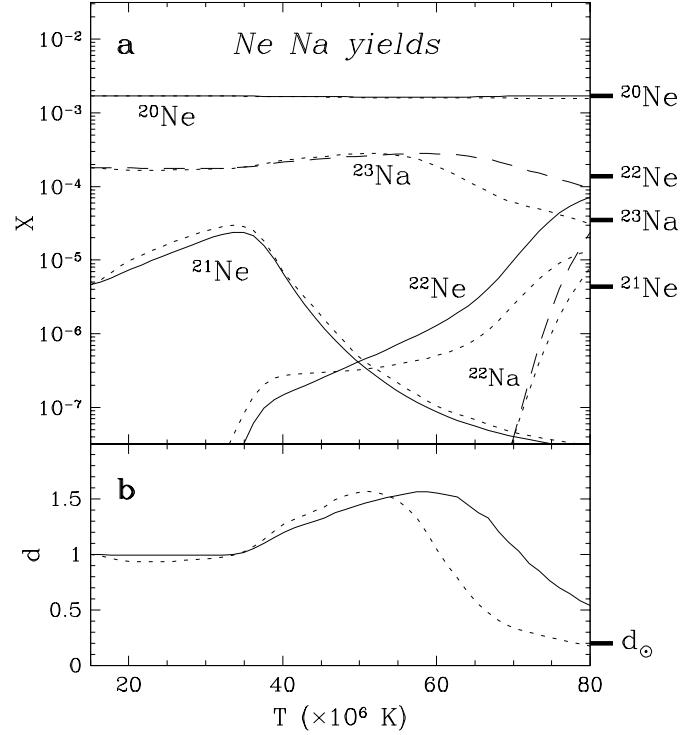}}
  \caption{{\bf a} Same as Fig.~\ref{Fig:CNOyields}, but for the stable nuclides
           involved in the NeNa chain.
           {\bf b} $d \equiv X(\mchem{Na}{23})/[X_0(\mchem{Na}{23})+\frac{23}{22}X_0(\mchem{Ne}{22})]$,
           where $X_0$ designates the initial abundance. The solar value $d_\odot$
           is indicated by the horizontal thick line at the right of the figure.
          }
  \label{Fig:NeNayields}
\end{figure}

The NeNa yields at hydrogen exhaustion are displayed in Fig.~\ref{Fig:NeNayields}a
(solid lines).
The main features concerning \chem{Na}{23} production are the following
(see Arnould et al. \cite{Arnould_etal95} for more details):

\vskip 1mm
\noindent - \chem{Ne}{22} is entirely transformed into \chem{Na}{23} at temperatures
as low as $15\times 10^6$~K. Consequently, the ratio $d \equiv
X(\mchem{Na}{23})/[X_0(\mchem{Na}{23})+\frac{23}{22}X_0(\mchem{Ne}{22})]$, where
$X_0$ designates the initial (here solar) abundance, is equal to one up to $\sim
35\times 10^6$~K (solid line in Fig.~\ref{Fig:NeNayields}b).

\vskip 1mm
\noindent - an extra production of \chem{Na}{23} is predicted at temperatures
exceeding $35\times 10^6$~K, reaching $\sim 60$\% at $T\simeq 60\times 10^6$~K
(Fig.~\ref{Fig:NeNayields}b).  This extra production results from the slight
burning of the abundant \chem{Ne}{20};

\vskip 1mm
\noindent - \chem{Na}{23} begins to burn by proton capture at
$T>60\times 10^6$~K, with $d=0.5$ at $T=80\times 10^6$~K.

 It is interesting to
further note that the El Eid \& Champagne (\cite{ElEid_Champagne95}) predict
\reac{Na}{23}{p}{\alpha}{Ne}{20} reaction rates always faster than those of
\reac{Na}{23}{p}{\gamma}{Mg}{24}, which ensures that the NeNa chain is a cycle.

  The impact of the new NACRE reaction rates on the NeNa yields
is shown by the dotted lines in Fig.~\ref{Fig:NeNayields}.
The \chem{Na}{23} destruction by p-capture is seen to be
much more important with NACRE than with the El Eid \& Champagne reaction rates above
$55\times 10^6$~K leading to about three times less \chem{Na}{23} abundances at temperatures
above $65\times 10^6$~K. Yet, the \chem{Na}{23} abundance at hydrogen exhaustion
does not decrease below its initial abundance at all temperatures of relevance for
our study (i.e. $d\ge d_\odot$ for $T<80\times 10^6$~K, see
Fig.~\ref{Fig:NeNayields}b). Arnould et al.~(\cite{Arnould_etal99}) further note that
the cycling character of the NeNa chain is put into question with the new NACRE
rates, but still depends on the large uncertainties affecting the p-capture rates
on \chem{Na}{23}.

It must be stressed that the large uncertainties affecting those
p-capture rates render the \chem{Na}{23} yields from the NeNa chain highly uncertain.
Adopting the most favorable NACRE
rates, Arnould et al.~(\cite{Arnould_etal99}) show, for example, that the \chem{Na}{23} mass
fraction at H exhaustion can still be $10^{-4}$ at $80\times 10^6$~K, i.e.
similar to the results obtained with the El Eid \& Champagne rates.

\section*{Appendix B: Analytical study of the surface \chem{Na}{23} abundance evolution
                      in AGB stars}

In order to analytically derive the evolution of the surface \chem{Na}{23} abundance
as a function of the surface C+N+O abundance, we make the following simplifying
assumptions:

\vskip 1mm
\noindent (a) the intershell \chem{C}{12} mass fraction $^{12}\!X_D$ left
over by the pulses is constant from one pulse to the next;

\vskip 1mm
\noindent (b) the dilution factor $f\equiv M_{dup}/M_{env}$ suffered by the
intershell nuclei when dredged-up into the envelope during a 3DUP
episode is constant from one 3DUP to the next;

\vskip 1mm
\noindent (c) the dilution factor $h\equiv M_B/M_D$ suffered by the intershell
\chem{N}{14} when injected from zone $B$ (Fig.~\ref{Fig:phases}) into the next
pulse is constant from one pulse to the next;

\vskip 1mm
\noindent (d) all \chem{C}{12} and \chem{O}{16} are transformed into
\chem{N}{14} in the CNO cycles, and all \chem{Ne}{22} into \chem{Na}{23}
in the NeNa chain. The extra-production of \chem{Na}{23} from
\chem{Ne}{20} is neglected, and the \chem{Na}{23} destruction by
p-capture is taken into account through the parameter
$d \equiv X(\mchem{Na}{23})/[X_0(\mchem{Na}{23})+\frac{23}{22}X_0(\mchem{Ne}{22})]$,
where $X_0$ designates the initial abundance. The parameter is assumed constant
from one pulse to the next. A value of $h=1$ means no \chem{Na}{23} destruction,
while a destruction of \chem{Na}{23} by p-captures leads to $h<1$;

\vskip 1mm
\noindent (e) the intershell \chem{N}{14} is assumed to be entirely
transformed into \chem{Ne}{22} in the pulses. The \chem{Ne}{22} destruction
by $\alpha$-capture in the hottest pulses is neglected (cf.
Sect.~\ref{Sect:synthetic calculations}).

\vskip 1mm
\noindent The validity of assumptions (a)-(d) is discussed in
Sect.~\ref{Sect:Na23discussion}, and that of (e) in Sect.~\ref{Sect:Ne22}.
The sensitivity of the \chem{Na}{23} abundance predictions to
$^{22}\!X_D$, $f$, $h$ and $d$, on the other hand, is analyzed at the end
of this Appendix.

Let us denote the surface mass fractions of \chem{C}{12}, \chem{N}{14}, \chem{O}{16},
\chem{Ne}{22} and \chem{Na}{23} by $c, n, o, e$ and $s$, respectively (i.e.
$c\equiv {^{12}\!X_s}$, ...$\;$, $s\equiv {^{23}\!X_s}$). Let us further denote
by $y$ the total abundance ${^{C\!N\!O}Y_s}$ of the CNO elements in
the envelope (cf. Sect.~\ref{Sect:Na23predictions}). The evolution of
the surface \chem{C}{12}, \chem{Ne}{22} and \chem{Na}{23} mass fractions, which are
dictated by the 3DUP episodes, are respectively given by
\begin{eqnarray}
\label{Eq:ci}
  c_i & = & c_{i-1} + f \; {^{12}\!X_D}\\
\label{Eq:ei0}
  e_i & = & e_{i-1} + f \; {^{22}\!X_{D,i}}\\
\label{Eq:si0}
  s_i & = & s_{i-1} + d \; f \; {^{23}\!X_{D,i}}
\end{eqnarray}
where the indice $i$ ($i-1$) refers to the abundances during interpulse
$i$ ($i-1$), and $^{22}\!X_{D,i}$ and $^{23}\!X_{D,i}$ are respectively the intershell
abundances of \chem{Ne}{22} and \chem{Na}{23} in zone $D$ after pulse $i$.
$^{22}\!X_{D,i}$ can be evaluated by recalling that the intershell
\chem{Ne}{22} abundance results
from the transformation of the CNO elements into \chem{N}{14} in the
HBS, followed by the mixing of \chem{N}{14} into the next pulse
with a dilution factor $h$, followed by its transformation in the pulse
into \chem{Ne}{22} by two $\alpha$-capture reactions. We thus have
\begin{displaymath}
  ^{22}\!X_{D,i}=h\times \frac{22}{14}\;y_{i-1},
\end{displaymath}
with $y_i\equiv \frac{14}{12}\,c_i+n_i+\frac{14}{16}\,o_i$.
Equation~(\ref{Eq:ei0}) then becomes
\begin{equation}
\label{Eq:ei}
 e_i = e_{i-1} + f \: h \; \frac{22}{14} \; y_{i-1}.
\end{equation}
We note that the evolution of $y_i$ is dictated by that of the surface carbon
abundance, since the surface nitrogen and oxygen abundances are not affected by the
3DUP episodes\footnote{
Actually oxygen is also produced in the pulses by \reac{C}{12}{\alpha}{\gamma}{O}{16}.
However, its mass fraction in the intershell layers is predicted by our standard models
to be about 20 times less than the intershell \chem{C}{12} mass fraction. For this
reason, we neglect in this analytical study its contribution to the surface CNO
abundance evolution.}.
We thus have
\begin{equation}
\label{Eq:yi0}
 y_i = \frac{14}{12}\;c_i \;+\; n_0 \;+\; \frac{14}{16}\;o_0
\end{equation}
with $c_i$ given by Eq.~(\ref{Eq:ci}), and $n_0$ and $o_0$ being respectively the surface mass
fractions of nitrogen and oxygen at the onset of the TP-AGB phase.

The intershell abundance $^{23}\!X_D$ of \chem{Na}{23}, on the other hand, results from the
conversion of \chem{Ne}{22} into \chem{Na}{23} by p-capture operating in the HBS.
Since \chem{Na}{23} is not destroyed in the pulses, we have\footnote{In reality, of
course, we have ${^{23}X_{D,i}} = d \times \left( s_{i-1}+\frac{23}{22}\;e_{i-1} \right)$.
For simplicity in the analytical derivations, however, we neglect the contribution of $s_{i-1}$
compared to that of $e_{i-1}$.  This approximation leads to a slight underestimation of $s_i$.}
\begin{displaymath}
  {^{23}X_{D,i}} = d \times \frac{23}{22}\;e_{i-1},
\end{displaymath}
and Eq.~(\ref{Eq:si0}) becomes
\begin{equation}
\label{Eq:si}
 s_i = s_{i-1} + f \; \frac{23}{22} \; e_{i-1}.
\end{equation}

  Equations (\ref{Eq:ci}), (\ref{Eq:ei}), (\ref{Eq:yi0}) and (\ref{Eq:si}) constitute a
system of recursive equations in the variables $c_i$ (or $y_i$), $e_i$ and $s_i$.
It can be solved rather simply as a function of pulse number $i$.  Knowing that
\begin{displaymath}
  \sum_{j=1}^{i-1}\; j\; =\; \frac{1}{2}\;i\;(i-1)
\end{displaymath}
and
\begin{displaymath}
  \sum_{j=1}^{i-1}\sum_{k=1}^{j-1}\; k\; =\; \frac{1}{6}\;i\;(i-1)\;(i-2),
\end{displaymath}
the solution to Eqs.~(\ref{Eq:ci}), (\ref{Eq:ei})-(\ref{Eq:si}) is
\begin{equation}
\label{Eq:cesi}
\begin{minipage}{6.8cm}
  \vskip -5mm
  \begin{eqnarray}
     c_i & = & c_0 + i\; f\; {^{12}\!X_D} \nonumber\\
     y_i & = & y_0 + i\; f\; {^{12}\!X_D} \nonumber\\
     e_i & = & e_0 + i\; f\; h\; \frac{22}{14}\; y_0
                   + \frac{i(i-1)}{2}\; f^2\; h\; \frac{22}{12}\;  {^{12}\!X_D} \nonumber\\
     s_i & = & s_0 + i\; d\; f\; \frac{23}{22}\; e_0
                   + \frac{i(i-1)}{2}\; d\; f^2\; h\; \frac{23}{14}\; y_0 \nonumber\\
         &   & \;\;\;\;\,+ \frac{i(i-1)(i-2)}{6}\; d\; f^3\; h\; \frac{23}{12} {^{12}\!X_D} \nonumber
  \end{eqnarray}
\end{minipage}
\end{equation}
where $c_0$, $y_0$, $e_0$ and $s_0$ are the surface abundances at the onset of the TP-AGB
phase.

  An analytical relation for the surface \chem{Na}{23} abundance as a function of
that of C+N+O can now easily be obtained by eliminating the
variable $i$ between Eqs. (\ref{Eq:cesi}) for $c_i$ and $s_i$. We
have, however, to be cautious that two consecutive pulses are necessary
before \chem{Na}{23} can build up from the envelope³s CNO elements. We
thus take the surface abundances $c_2$, $y_2$, $e_2$ and $s_2$ at the
second pulse as initial conditions in Eqs.~(\ref{Eq:cesi}) instead of the
surface abundances
at the onset of the TP-AGB phase [replace $c_0$, $y_0$, $e_0$, $s_0$ respectively
by $c_2$, $y_2$, $e_2$, $s_2$ in Eqs.~(\ref{Eq:cesi}],
and eliminate $i$ between the expressions
for $c_i$ and $s_i$ so obtained. This leads to a relation between
$s$ and $y$ valid from the second 3DUP on, i.e. for
$y_i \ge y_0 + 2\; f\; {^{12}\!X_D}$. The surface \chem{Na}{23}
abundance after the first 3DUP (i.e. at $y = y_0 + f\; {^{12}\!X_D}$),
on the other hand, is immediately obtained from Eqs.~(\ref{Eq:cesi}) by
taking $i=1$. The result is, after some elementary algebraic
manipulations,
\begin{displaymath}
\left\{ \;
\begin{minipage}{6.8cm}
\vskip -5mm
  \begin{eqnarray}
    s & = & s_0 + f\; \frac{23}{22}\; e_0
                          \mbox{\hspace*{25mm}for} \;  y = y_0 + f\; {^{12}\!X_D} \nonumber
  \end{eqnarray}
  \begin{eqnarray}
\label{Eq:c-s}
    s & = & d\; \left[ \frac{23}{72}\; h\; {^{12}\!X_D}\;
                       \left( \frac{y-y_2}{\frac{14}{12}\;{^{12}\!X_D}} \right)^{\!3} \right. \\
      &   & \;\;\;\;\;\; +\; \frac{1}{2}\; h \left( \frac{23}{14}\; y_2 - f\; \frac{23}{12}\; {^{12}\!X_D} \right)\!
            \left( \frac{y-y_2}{\frac{14}{12}\;{^{12}\!X_D}} \right)^{\!2} \nonumber\\
      &   & \;\;\;\;\;\; +\; \left( \frac{23}{22}\; e_2 - \frac{1}{2}\; f\; h\; \frac{23}{14}\; y_2
                       + \frac{1}{3}\; f^2\; h\; \frac{23}{12}\; {^{12}\!X_D} \right)\; \nonumber\\
      &   & \;\;\;\;\;\; \left. \;\;\;\;\;\;\;\;\times
            \left( \frac{y-y_2}{\frac{14}{12}\;{^{12}\!X_D}} \right) \; \right] \nonumber\\
      &   & + s_2,        \mbox{\hspace*{33mm}for} \; y \ge y_0 + 2\; f\; {^{12}\!X_D}, \nonumber
  \end{eqnarray}
\end{minipage}
\right.
\end{displaymath}
with [from Eqs.~(\ref{Eq:yi0}) and (\ref{Eq:cesi})]
\begin{equation}
\label{Eq:cesy2}
\begin{minipage}{6.8cm}
  \vskip -5mm
  \begin{eqnarray}
    y_2 & = & y_0 + 2\; f\; \frac{14}{12}\; {^{12}\!X_D} \nonumber\\
    e_2 & = & c_0 + 2\; f\; h\; \frac{22}{14}\; y_0 +  f^2\; h\; \frac{22}{12}\; {^{12}\!X_D} \nonumber\\
    s_2 & = & s_0 + 2\; d\; f\; \frac{23}{22}\; e_0 +  d\; f^2\; h\; \frac{23}{14}\; y_0. \nonumber
  \end{eqnarray}
\end{minipage}
\end{equation}

\begin{figure}
  \resizebox{\hsize}{!}{\includegraphics{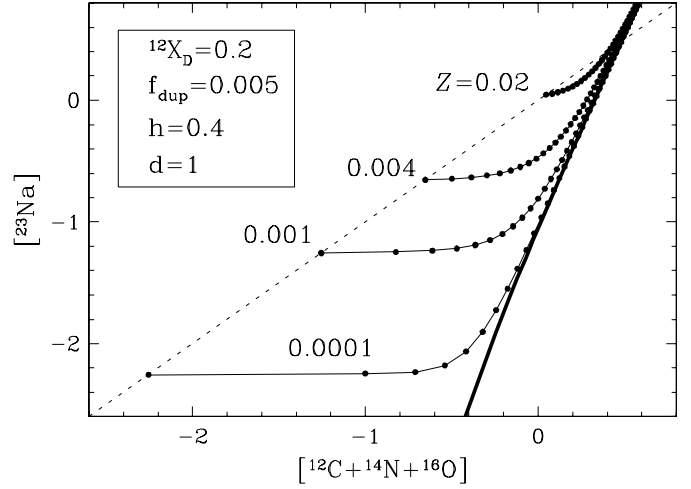}}
  \caption{Surface \chem{Na}{23} abundance, expressed in dex
           relative to the solar abundance \{i.e.
           $\log[X(\mchem{Na}{23})/X_\odot(\mchem{Na}{23})]$\}, predicted 
           analytically by Eq.~(\ref{Eq:c-s}) for different metallicities $Z$ as
           labeled in the figure. The predictions are given as a function
           of the surface CNO abundances, also expressed in dex (by number)
           relative to the solar abundance. The `canonical' values used for the parameters
           $^{12}\!X_D$, $f$ and $h$ entering (Eq.~\ref{Eq:c-s}) are given in the
           insert at the upper-left corner of the figure. The filled circles on
           each track locate the pulses
           as predicted by (Eq.~\ref{Eq:ci}). The dashed line
           represents the abundances of \chem{Na}{23} versus the CNO abundances
           if a solar distribution of elements heavier than helium is assumed at different
           metallicities. The thick line is the `line of primary \chem{Na}{23}
           production' defining the \chem{Na}{23} abundance evolution in zero
           metallicity stars (see text).
          }
  \label{Fig:analZ}
\end{figure}

\begin{figure}
  \resizebox{\hsize}{!}{\includegraphics{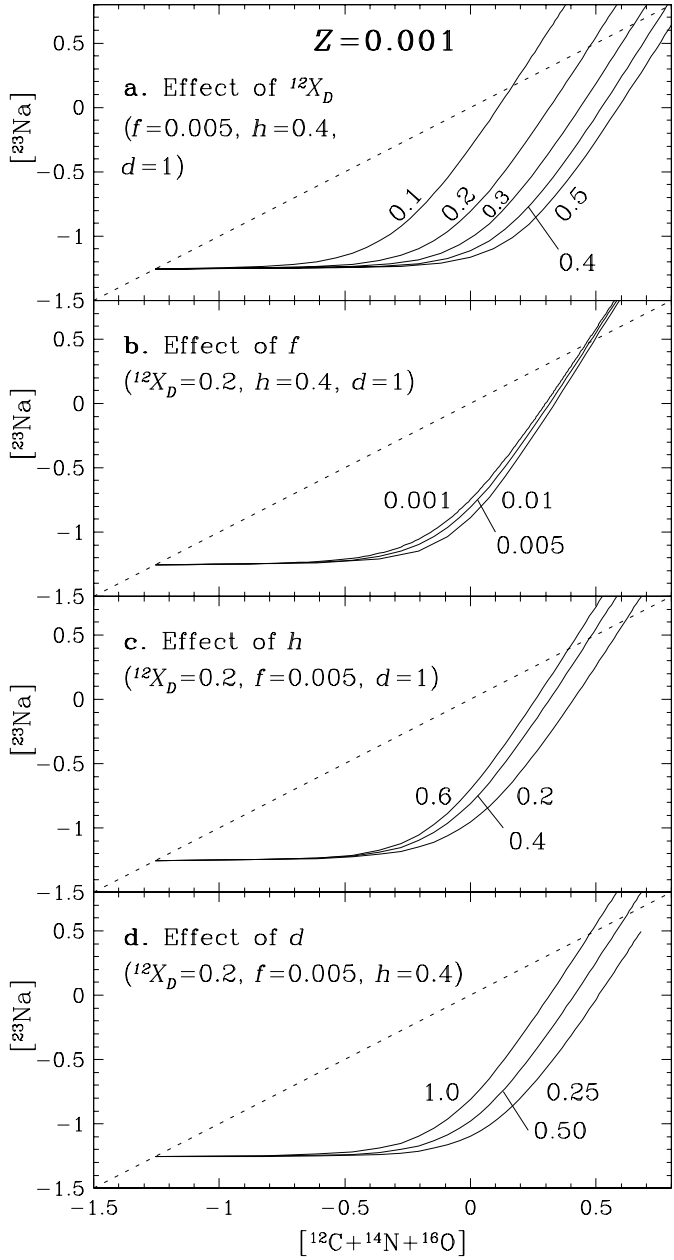}}
  \caption{Same as Fig.~\ref{Fig:analZ}, but
           {\bf a} for $^{12}\!X_D=0.1, 0.2, 0.3, 0.4$ and 0.5 at fixed $f$, $h$ and $d$,
           {\bf b} for $f=0.001, 0.005$ and 0.01 at fixed $^{12}\!X_D$, $h$ and $d$,
           {\bf c} for $h=0.2, 0.4$ and 0.6 at fixed $^{12}\!X_D$, $f$ and $d$,
           and {\bf d} for $d=1.0, 0.5$ and 0.25 at fixed $^{12}\!X_D$, $f$ and $h$.
           All tracks correspond to $Z=0.001$ initial composition.
          }
  \label{Fig:analXfhd}
\end{figure}

\subsection*{Zero metallicity stars}

Let us first consider the \chem{Na}{23} production in zero
metallicity stars\footnote{See footnote \ref{NOTE:Z=0}}
(i.e. the production of pure primary sodium). Putting
$c_0=n_0=o_0=e_0=s_0=0$ in Eqs.~(\ref{Eq:c-s}) and
(\ref{Eq:cesy2}), we get
\begin{eqnarray}
\label{Eq:c-s,Z=0}
  s & = & \frac{23}{12}\;d\;h\;{^{12}\!X_D} \nonumber\\
    &   & \times \left[ \frac{1}{6}\;\left(\frac{c-2\,f\,{^{12}\!X_D}}{{^{12}\!X_D}}\right)^3
                        +\frac{1}{2}\;f\;\left(\frac{c-2\,f\,{^{12}\!X_D}}{{^{12}\!X_D}}\right)^2
                 \right. \nonumber\\
    &   & \;\;\;\;\;\;\;
                 \left. +\frac{1}{3}\;f^2\;\left(\frac{c-2\,f\,{^{12}\!X_D}}{{^{12}\!X_D}}\right)
                 \right]
\end{eqnarray}

Equation (\ref{Eq:c-s,Z=0}) is shown in thick line in
Fig.~\ref{Fig:analZ} for `canonical' values of the parameters
${^{12}\!X_D}=0.2$, $f=0.005$, $h=0.4$ and $d=1$ (see
Sect.~\ref{Sect:Na23discussion}). It represents the abundance of primary
\chem{Na}{23} resulting from the sole contribution of the primary
\chem{C}{12} produced in the HeBS. It is therefore called
the {\it line of primary sodium enrichment} (LOPSE).

  The sensitivity of the LOPSE to the
parameters ${^{12}\!X_D}$, $f$, $h$ and $d$ can be estimated from
Eq.~(\ref{Eq:c-s,Z=0}). Knowing that $f<<1$ in most circumstances, we can
neglect the terms in $f$ and $f^2$ in the right hand side of
Eq.~(\ref{Eq:c-s,Z=0}), and write to first order
\begin{equation}
\label{Eq:c-s,Z=0,f=0}
  s = \frac{23}{72}\;d\;h\;\frac{c^3}{({^{12}\!X_D})^2}.
\end{equation}

Several important conclusions can immediately be drawn from
Eq.~(\ref{Eq:c-s,Z=0,f=0}):

\vskip 1mm
\noindent 1. the surface \chem{Na}{23} abundance increases as
the third power of the surface \chem{C}{12} abundance. In other words,
the LOPSE is represented, to first order, by a straight
line in the [C+N+O]$-$[\chem{Na}{23}] plane with a slope of 3
(thick solid line in Fig.~\ref{Fig:analZ});

\vskip 1mm
\noindent 2. the LOPSE is insensitive, to first order, to $f$ [that
parameter does not appear in Eq.~(\ref{Eq:c-s,Z=0,f=0})].
In other words, {\it primary $^{23}\!Na$ abundance prediction as a function of
primary \chem{C}{12} abundance is insensitive to the dredge-up efficiency};

\vskip 1mm
\noindent 3. the value of $^{23}\!X_s$ predicted by the LOPSE is linearly
proportional to $d$, as expected since that parameter represents the `destruction'
factor of \chem{Na}{23} by p-capture;

\vskip 1mm
\noindent 4. the value of $^{23}\!X_s$ is also linearly
proportional to the parameter $h$ at given surface
\chem{C}{12} abundance. Indeed, if the dilution of the intershell \chem{N}{14}
into the pulse is low (i.e. the dilution factor $h$ is high) then the abundance
of \chem{Ne}{22} emerging from the pulse is high, leading to higher
\chem{Na}{23} production during the next interpulse. Thus, high $h$
factors imply high $^{23}\!X_s$ abundances at a given \chem{C}{12} abundance,
as expressed by Eq.~(\ref{Eq:c-s,Z=0,f=0});

\vskip 1mm
\noindent 5. the most influential parameter on the LOPSE is ${^{12}\!X_D}$,
which appears to the second power in Eq.~(\ref{Eq:c-s,Z=0,f=0}).
This results from the fact that the surface
\chem{C}{12} abundance $^{12}\!X_s$ increases linearly pulse after pulse,
the $^{12}\!X_s$ increment being proportional to ${^{12}\!X_D}$, while
$^{23}\!X_s$ increases to the third power of the pulse number. The
relation is such that $^{23}\!X_s$ increases with decreasing
${^{12}\!X_D}$ at a given surface \chem{C}{12} abundance\footnote{The fact that
$^{23}\!X_s$ decreases with increasing ${^{12}\!X_D}$ seems against intuition,
since \chem{Na}{23} is synthesized from the intershell \chem{C}{12}. What actually
happens is that the surface \chem{C}{12}, and thus [C+N+O], increases more rapidly
with ${^{12}\!X_D}$, shifting the tracks in the [C+N+O]--[Na] diagram to the
right.}

\subsection*{Non-zero metallicity stars}

The evolution of $^{23}\!X_s$ at $Z\neq 0$ is dictated by
Eqs.~(\ref{Eq:c-s}), and shown in thin solid lines in Fig.~\ref{Fig:analZ}
for several metallicities (the initial abundances are scaled from solar
to the required metallicity as represented by the dotted
line in Fig.~\ref{Fig:analZ}). The surface
\chem{Na}{23} abundance stays constant until the surface
C+N+O abundance reaches the LOPSE.
This results from
the fact that the abundance of primary \chem{Na}{23} produced during the
first pulses (i.e. when the C+N+O abundance is still far below the value
required by the LOPSE) is much lower than the initial $^{23}\!X_s$.
No variation of the surface \chem{Na}{23}
abundance is thus expected. Only when the surface C+N+O abundance
approaches the LOPSE does the increasing contribution of primary
\chem{Na}{23} become competitive with $^{23}\!X_s$.
From then on, the surface \chem{Na}{23} abundance is
mainly provided by the primary \chem{Na}{23}, and $^{23}\!X_s$ increases
asymptotically towards the LOPSE.
Those conclusions confirm the ones obtained from the synthetic
predictions presented in Sect.~\ref{Sect:synthetic calculations}.

\subsection*{Sensitivity to parameters}

\paragraph{Sensitivity to ${^{12}\!X_D}$}

The sensitivity of $^{23}\!X_s$ predictions to
${^{12}\!X_D}$ is illustrated in Fig.~\ref{Fig:analXfhd}a.
As already discussed in the analysis
of zero metallicity stars (see above), ${^{12}\!X_D}$ has
a main influence on the location of the LOPSE. As a consequence, it
affects directly the evolution of $^{23}\!X_s$ at all metallicities.
The lines at different
${^{12}\!X_D}$ values are practically parallel to each other, the ones
with the lowest ${^{12}\!X_D}$ predicting the highest surface
\chem{Na}{23} abundances. At a given [C+N+O] value, the increase in
[\chem{Na}{23}] on the LOPSE is twice the decrease in ${^{12}\!X_D}$
expressed in dex [$^{23}\!X_s$ is inversely proportional to the square
of ${^{12}\!X_D}$ in Eq.~(\ref{Eq:c-s,Z=0,f=0})].

\paragraph{Sensitivity to $f$}

  Figure~\ref{Fig:analXfhd}b reveals that $^{23}\!X_s$ is almost
insensitive to $f$. This property is expected from the independence of the LOPSE
on that parameter [see Eq.~(\ref{Eq:c-s,Z=0,f=0})].

\paragraph{Sensitivity to $h$}

  The effect of $h$ on $^{23}\!X_s$ predictions is
illustrated in Fig.~\ref{Fig:analXfhd}c. The importance of that parameter results directly from
the linear dependence of $^{23}\!X_s$ on $h$ on the LOPSE.
At a given [C+N+O] value, the increase in [\chem{Na}{23}] on the LOPSE is
equal to that of $h$ expressed in dex [$^{23}\!X_s$ is linearly proportional
to $h$ in Eq.~(\ref{Eq:c-s,Z=0,f=0})]. We further note that the value of [C+N+O]
from which $^{23}\!X_s$ begins to increase is independent of $h$,
contrary to the case of ${^{12}\!X_D}$ (cf. Figs.~\ref{Fig:analXfhd}a and c).

\paragraph{Sensitivity to $f$}

  The effect of $d$ on $^{23}\!X_s$ is similar to the effect of $h$, both of those
parameters influencing linearly the LOPSE. Its effect is illustrated in
Fig.~\ref{Fig:analXfhd}d.

\vskip 3mm
\noindent {\small{\it Acknowledgments.}
I thank Dr. Hans Van Winckel for his careful reading and commenting on the manuscript
and Dr. G. Meynet for many useful discussions.

\end{document}